\documentclass[aps,twocolumn,showpacs]{revtex4}
\usepackage{amsmath}
\usepackage{epsfig}
\usepackage{multirow}
\usepackage{appendix}
\usepackage{amssymb}
\usepackage{epstopdf}

\begin{document}
	
\title{Investigating $\Xi$ resonances from pentaquark perspective}
\author{Ye Yan$^1$}\email{221001005@njnu.edu.cn}
\author{Qi Huang$^1$}\email{06289@njnu.edu.cn}
\author{Xinmei Zhu$^2$}\email{xmzhu@yzu.edu.cn(Corresponding author)}
\author{Hongxia Huang$^1$}\email{hxhuang@njnu.edu.cn(Corresponding author)}
\author{Jialun Ping$^1$}\email{jlping@njnu.edu.cn}
\affiliation{$^1$School of Physics and Technology, Nanjing Normal University, Nanjing 210097, People's Republic of China}
\affiliation{$^2$Department of Physics, Yangzhou University, Yangzhou 225009, People's Republic of China}

\begin{abstract}
We have investigated the $qss\bar{q}q$ ($q = u$ or $d$) system to find possible pentaquark explanations for the $\Xi$ resonances.
The bound state calculation is carried out within the framework of the quark delocalization color screening model.
The scattering processes are also studied to examine the possible resonance states.
The current results indicate that the $\Xi(1950)$ can be interpreted as $\Lambda \bar{K}^*$ state with $J^P = 1/2^-$.
Three states are identified that match the $\Xi(2250)$, which are $\Sigma^* \bar{K}^*$ state with $J^P = 3/2^-$, $\Sigma^* \bar{K}^*$ state with $J^P =5/2^-$, and $\Xi^* \rho$ state with $J^P =5/2^-$.
This may explain the conflicting experimental values for the width of the $\Xi(2250)$.
A new $\Xi$ resonance is predicted, whose mass and width are 2066--2079 MeV and 186--189 MeV, respectively.
These results contribute to understanding the nature of the $\Xi$ resonances and to the future search for new $\Xi$ resonances.
Moreover, it is meaningful to further investigate the $\Xi$ resonances from an unquenched picture on the basis of pentaquark investigation.
\end{abstract}
	
	
\maketitle

\setcounter{totalnumber}{5}
	
\section{Introduction}
The study of strange baryons forms a bridge between light-flavor baryons and heavy-flavor baryons, playing a critical role in our comprehension of the baryon spectrum.
Through the efforts of experimental collaborations and the accumulation of experimental data, the information about strange baryons has been continually unveiled~\cite{SIDDHARTA:2011dsy,CLAS:2013rjt,CLAS:2014tbc,BESIII:2015dvj,Belle:2018mqs,Belle:2019zco,Belle:2018lws,BESIII:2019cuv,Belle:2021gtf,LHCb:2020jpq,BGOOD:2021sog,ALICE:2022yyh,J-PARCE31:2022plu,Amsterdam-CERN-Nijmegen-Oxford:1977bvi}.
However, compared with $\Lambda$ and $\Sigma$ baryons in the $S = -1$ sector, the information regarding $\Xi$ baryons in the $S = -2$ sector is even more obscure.
So far, only the ground octet and decuplet states with four-star ratings, $\Xi(1320)$ with $J^P = 1/2^+$ and $\Xi(1530)$ with $J^P = 3/2^+$, as well as the three-star-rated  $\Xi(1820)$ with $J^P = 3/2^-$, have determined spin-parity quantum numbers~\cite{Workman:2022ynf}.
Our understanding of the remaining eight $\Xi$ resonances is still insufficient.

Understanding these $\Xi$ resonances and explaining their properties and structures has always been a focal point of theoretical work, helping us gain a better comprehension of strong interactions and QCD (quantum chromodynamics).
The most direct approach is to understand these states from the perspective of traditional three-quark configuration.
The $\Xi$ spectrum has been investigated as three-quark configuration in the framework of various quark models~\cite{Chao:1980em,Capstick:1986ter,Glozman:1995fu,Glozman:1997ag,Valcarce:2005rr,Bijker:2000gq,Arifi:2022ntc,Pervin:2007wa,Xiao:2013xi,Menapara:2021dzi},
Skyrme model~\cite{Oh:2007cr},
lattice QCD~\cite{Engel:2013ig},
QCD sum rules~\cite{Jido:1996zw,Lee:2002jb,Aliev:2018hre},
and large-$N_c$ analysis~\cite{Semay:2007cv,Schat:2001xr,Goity:2003ab,Matagne:2004pm,Matagne:2006zf}.
According to the numerical results, the interpretation of $\Xi(1318)$, $\Xi(1530)$, $\Xi(1820)$, and $\Xi(1950)$ as three-quark excited states is consistent with the $\Xi$ baryon spectrum obtained by most of the aforementioned approaches.
In addition, interpreting $\Xi(1620)$ as a $J^P = 1/2^-$ state is supported in Ref.~\cite{Oh:2007cr} using the Skyrme model and Ref.~\cite{Jido:1996zw} using the QCD sum rules.
In Ref.~\cite{Arifi:2022ntc}, the spin-parity quantum numbers of $J^P = 1/2^-$ is also favored for $\Xi(1620)$ by studying its property of strong decay.
As for $\Xi(1690)$, in the Skyrme model~\cite{Oh:2007cr}, it is predicted to have $J^P = 1/2^-$.
The same $J^P = 1/2^-$ prediction for $\Xi(1690)$ can be found in other theoretical works, by using different quark models~\cite{Arifi:2022ntc,Pervin:2007wa,Xiao:2013xi} and the QCD sum rules~\cite{Aliev:2018hre}.
A $J^P = 1/2^+$ $\Xi$ state derived from the nonrelativistic quark model also matches well with $\Xi(1690)$~\cite{Chao:1980em}.
In Ref.~\cite{Xiao:2013xi}, $\Xi(1950)$ might correspond to several different $\Xi$ resonances, and $\Xi(2030)$ seems to have $J^P = 3/2^+$, although this conflicts with the spin quantum number provided by moment analysis~\cite{Amsterdam-CERN-Nijmegen-Oxford:1977bvi}.
In Ref.~\cite{PavonValderrama:2011gp} analysis, based on the experimental measurements, it is proposed that $\Xi(1950)$ might correspond to three $\Xi$ resonances, which are $J^P = 1/2^-$ decuplet, $J^P = 5/2^+$ octet and $J^P = 5/2^-$ octet.

In recent years, many states have been observed by experimental collaborations that are difficult to explain using traditional hadronic states.
In particular, the $P_c$ and $P_{cs}$ states have been widely regarded as promising candidates for pentaquarks in researches~\cite{LHCb:2020jpq,LHCbPc2015,LHCbPc2019,LHCb:2022ogu,Chen:2016qju,Liu:2019zoy,Yang:2020atz}.
Additionally, the $\Lambda(1405)$ has also attracted extensive attention~\cite{CLAS:2013rjt,CLAS:2014tbc,BGOOD:2021sog,ALICE:2022yyh,J-PARCE31:2022plu}.
While the pole position and the structure of the $\Lambda(1405)$ are still in controversy, many work support its two poles nature and the possible molecular explanation~\cite{Jido:2003cb,Hyodo:2011ur,Meissner:2020khl,Hyodo:2020czb,Qin:2020gxr,Mai:2020ltx,BaryonScatteringBaSc:2023zvt}.
The progress inspires us to explore $\Xi$ resonances from a broader perspective, suggesting possibilities beyond the traditional three-quark configuration, including pentaquark configuration, dynamical effect, or unquenched picture.

Some theoretical studies have delved into the interpretation of $\Xi$ resonances beyond the conventional three-quark configuration.
In the framework of the one-boson-exchange model, the $\Xi(1620)$ can be explained as a $\bar{K} \Lambda$ molecular state, and a $\bar{K} \Sigma$ molecular state is also predicted~\cite{Chen:2019uvv}.
Using the same model, it is suggested that $\Xi(2030)$ can be assigned as a $P$-wave $\bar{K}\Sigma / \rho \Xi / \bar{K} \Lambda / \phi \Xi / \omega \Xi$ molecular state with $J^P = 5/2^+$~\cite{Hei:2023eqz}.
In Ref.~\cite{Sarti:2023wlg}, femtoscopic data is used to constrain the parameters of a low-energy effective QCD Lagrangian and the primary composition of the $\Xi(1620)$ is suggested to be $\eta \Xi$.
Additionally, the decays and the production of the $\Xi(1620)$ are studied in the framework of the effective Lagrangian approach.
The radiative decays of the $\Xi(1620)$ is studied by assuming that it is a $\Lambda \bar{K}-\Sigma \bar{K}$ molecular state in Ref.~\cite{Huang:2020taj}.
In Ref.~\cite{Guo:2023gvx}, the production of the $\Xi(1620)$ in the $K^- p$ scattering process is investigated, where the $\Xi(1620)$ is considered as a $\bar{K} \Lambda$ molecular state.
The production of the $\Xi(1690)$ from the $K^- p \rightarrow K^+ K^- \Lambda$ reaction is also studied within the effective Lagrangian approach in Ref.~\cite{Ahn:2018hbc}
Based on the Bethe-Salpeter equation approach~\cite{Garcia-Recio:2003ejq}, it is suggested that the quantum numbers $J^P = 1/2^-$ be assigned to the $\Xi(1620)$ and the $\Xi(1690)$.
Using a similar approach~\cite{Wang:2019krq}, the $\Xi(1620)$ can be explained as $\Lambda \bar{K}$ and $\Sigma \bar{K}$ bound states with $J^P = 1/2^-$, respectively.
In Ref.~\cite{Feijoo:2023wua}, the $\Xi(1620)$ and the $\Xi(1690)$ are studied as molecular states from $S = -2 $ meson-baryon interaction using an extended Unitarized Chiral Perturbation Theory.
In Ref.~\cite{Shevchenko:2015oea}, $\Xi(1950)$ is explained as a $\bar{K} \bar{K} N$ state using Faddeev-type AGS equations.

By constructing the dynamical baryon-meson scattering processes in the strangeness $S = -2$ sector, many results have been achieved in the framework of the chiral unitary approach.
According to the findings presented in Refs.~\cite{Ramos:2002xh,Ramos:2003mu,Nishibuchi:2023acl,Miyahara:2016yyh,Li:2023olv,Gamermann:2011mq}, signals associated with the $\Xi(1620)$ are dynamically generated, and the spin-parity quantum numbers of the $\Xi(1620)$ is identified to be $J^P = 1/2^-$.
Regarding the nature of this resonance, in Refs.~\cite{Ramos:2002xh,Miyahara:2016yyh,Li:2023olv}, results support that $\Xi(1620)$ couples strongly to the $\Xi \pi$ and the $\Lambda \bar{K}$ channels.
In Ref.~\cite{Gamermann:2011mq}, $\Xi(1620)$ is considered to strongly couple to the $\Xi \pi$ channel.
It is demonstrated in Ref.~\cite{Nishibuchi:2023acl} that the spectrum of the $\Xi(1620)$ is distorted by the effect of the nearby $\Lambda \bar{K}$ threshold.
In Ref.~\cite{Sun:2021bgl}, a $J^P = 1/2^-$ resonance state with a mass of about 1550 MeV and a decay width of 120--200 MeV is dynamically generated, which is inconsistent with the $\Xi(2120)$.
Meanwhile, in the similar framework~\cite{Miyahara:2016yyh,Li:2023olv,Liu:2023jwo,Sekihara:2015qqa,Gamermann:2011mq,Khemchandani:2016ftn} , a $\Xi$ state with $J^P = 1/2^-$ can be dynamically generated near the $\Sigma \bar{K}$ threshold, which can be identified with the $\Xi(1690)$ resonance.
In Refs.~\cite{Gamermann:2011mq,Sarkar:2004jh}, $\Xi(1820)$ with $J^P = 3/2^-$ can be dynamically generated in the meson-baryon scattering process using the coupled-channel unitary approach.
A similar conclusion can also be found in Ref.~\cite{Kolomeitsev:2003kt}, within the $\chi$-BS(3) approach.
While in Ref.~\cite{Molina:2023uko}, chiral unitary approach for baryon-meson interaction gives rise to a narrow state and a wide state around the $\Xi(1820)$, indicating that the $\Xi(1820)$ may actually be two states.
For other $\Xi$ resonances, $\Xi(1950)$ is identified with a spin-parity $J^P = 1/2^-$ state~\cite{Gamermann:2011mq}.
A narrow $J^P = 3/2^-$ state with mass $\sim$2050 MeV is found, and related to the $\Xi(2120)$ in Ref.~\cite{Khemchandani:2016ftn}.
In Ref.~\cite{Oset:2010tof}, two states are generated that can be associated to the $\Xi(1950)$ and the $\Xi(2120)$, using the formalism that produces doublets of degenerate $J^P = 1/2^-, 3/2^-$ states.

Studying hadron-hadron scattering processes is crucial for understanding resonances.
In addition to the above theoretical methods, the quark delocalization color screening model (QDCSM) provides an alternative approach for studying multiquark systems.
It offers a robust description of the $NN$ and $YN$ interactions and the properties of the deuteron~\cite{Ping:2000dx,Ping:1998si,Wu:1998wu,Pang:2001xx}.
Moreover, building on the foundation of the Kohn-Hulth\'{e}n-Kato (KHK) variational method~\cite{Oka:1981ri}, the model is employed to compute hadron-hadron scattering phase shifts and investigate exotic hadronic states~\cite{Huang:2015uda,Huang:2018wed,Huang:2023jec}.
In Ref.~\cite{Huang:2018wed}, hidden-charm pentaquark resonances are investigated in the hadron-hadron scattering process, and three of the obtained states are consistent with the $P_c$ states later reported by the LHCb collaboration in 2019.

Using this model, we have investigated the pentaquark interpretations of the $\Lambda_c$ baryons~\cite{Yan:2022nxp} and $\Omega_c$ baryons~\cite{Yan:2023tvl}, successfully obtaining masses and widths that agree with experimental values.
Therefore, we extend the application of this model to investigate the $qss\bar{q}q$ system, aiming to explore potential molecular interpretations of $\Xi$ resonances.
In this work, we investigate the energy spectrum of the pentaquark system.
The scattering processes are studied to assess the presence of resonance states and provide information about the characteristics of the identified states.

This paper is organized as follows.
In the next section, we provide a detailed overview of QDCSM, scattering process, and the construction of wave function.
In Sec. III, numerical results and discussions are given, followed by a summary in Sec. IV.

\section{THEORETICAL FORMALISM}

\subsection{Quark delocalization color screening model}
Herein, the QDCSM is employed to investigate the properties of $qss\bar{q}q$ system.
The QDCSM is an extension of the native quark cluster model~\cite{DeRujula:1975qlm,Isgur:1978xj,Isgur:1978wd,Isgur:1979be} and has been developed to address multiquark systems~\cite{Ping:2000cb,Huang:2013nba,Xue:2020vtq,Yan:2023tvl}.
In this section, we mainly introduce the salient features of this model.

The general form of the pentaquark Hamiltonian is given by
	\begin{align}
		H=&\sum_{i=1}^5\left(m_i+\frac{\boldsymbol{p}_{i}^{2}}{2m_i}\right)-T_{\mathrm{c} . \mathrm{m}} +\sum_{j>i=1}^5 V(\boldsymbol{r}_{ij}),
	\end{align}
where $m_i$ is the quark mass, $\boldsymbol{p}_{i}$ is the momentum of the quark, and $T_{\mathrm{c.m.}}$ is the center-of-mass kinetic energy.
The dynamics of the pentaquark system is driven by a two-body potential
	\begin{align}
		V(\boldsymbol{r}_{ij})= & V_{\mathrm{CON}}(\boldsymbol{r}_{ij})+V_{\mathrm{OGE}}(\boldsymbol{r}_{ij})+V_{\chi}(\boldsymbol{r}_{ij}).
	\end{align}
The most relevant features of QCD at its low energy regime---color confinement ($V_{\mathrm{CON}}$), perturbative one-gluon exchange interaction ($V_{\mathrm{OGE}}$), and dynamical chiral symmetry breaking ($V_{\chi}$)---have been taken into consideration.

Here, a phenomenological color screening confinement potential ($V_{\mathrm{CON}}$) is used as
\begin{align}
	V_{\mathrm{CON}}(\boldsymbol{r}_{ij}) = & -a_{c}\boldsymbol{\lambda}_{i}^{c} \cdot \boldsymbol{\lambda}_{j}^{c}\left[  f(\boldsymbol{r}_{ij})+V_{0}\right],
\end{align}
\begin{align}
	f(\boldsymbol{r}_{ij}) =& \left\{\begin{array}{l}
		\boldsymbol{r}_{i j}^{2}, ~~~~~~~~~~~~~ ~i,j ~\text {occur in the same cluster } \\
		\frac{1-e^{-\mu_{q_{i}q_{j}} \boldsymbol{r}_{i j}^{2}}}{\mu_{q_{i}q_{j}}},  ~~~i,j ~\text {occur in different cluster }
	\end{array}\right.   \nonumber
\end{align}
where $a_c$, $V_{0}$ and $\mu_{q_{i}q_{j}}$ are model parameters, and $\boldsymbol{\lambda}^{c}$ stands for the SU(3) color Gell-Mann matrices.
Among them, the color screening parameter $\mu_{q_{i}q_{j}}$ is determined by fitting the deuteron properties, nucleon-nucleon scattering phase shifts, and hyperon-nucleon scattering phase shifts, respectively, with $\mu_{qq}=0.45$, $\mu_{qs}=0.19$, and $\mu_{ss}=0.08~$fm$^{-2}$, satisfying the relation---$\mu_{qs}^{2}=\mu_{qq}\mu_{ss}$~\cite{ChenM}.

In the present work, we mainly focus on the low-lying negative parity $qss\bar{q}q$ pentaquark states of the $S$-wave, so the spin-orbit and tensor interactions are not included.
The one-gluon exchange potential ($V_{\mathrm{OGE}}$), which includes Coulomb and chromomagnetic interactions, is written as
	\begin{align}
		V_{\mathrm{OGE}}(\boldsymbol{r}_{ij})= &\frac{1}{4}\alpha_{s_{q_i q_j}} \boldsymbol{\lambda}_{i}^{c} \cdot \boldsymbol{\lambda}_{j}^{c}  \\
		&\cdot \left[\frac{1}{r_{i j}}-\frac{\pi}{2} \delta\left(\mathbf{r}_{i j}\right)\left(\frac{1}{m_{i}^{2}}+\frac{1}{m_{j}^{2}}+\frac{4 \boldsymbol{\sigma}_{i} \cdot \boldsymbol{\sigma}_{j}}{3 m_{i} m_{j}}\right)\right],   \nonumber \label{Voge}
	\end{align}
where $\boldsymbol{\sigma}$ is the Pauli matrices and $\alpha_{s_{q_i q_j}}$ is the quark-gluon coupling constant.

The dynamical breaking of chiral symmetry results in the SU(3) Goldstone boson exchange interactions appear between constituent light quarks $u, d$, and $s$.
Hence, the chiral interaction is expressed as
\begin{align}
	V_{\chi}(\boldsymbol{r}_{ij})= & V_{\pi}(\boldsymbol{r}_{ij})+V_{K}(\boldsymbol{r}_{ij})+V_{\eta}(\boldsymbol{r}_{ij}).
\end{align}
Among them
\begin{align}
V_{\pi}\left(\boldsymbol{r}_{i j}\right) =&\frac{g_{c h}^{2}}{4 \pi} \frac{m_{\pi}^{2}}{12 m_{i} m_{j}} \frac{\Lambda_{\pi}^{2}}{\Lambda_{\pi}^{2}-m_{\pi}^{2}} m_{\pi}\left[Y\left(m_{\pi} \boldsymbol{r}_{i j}\right)\right. \nonumber \\
&\left.-\frac{\Lambda_{\pi}^{3}}{m_{\pi}^{3}} Y\left(\Lambda_{\pi} \boldsymbol{r}_{i j}\right)\right]\left(\boldsymbol{\sigma}_{i} \cdot \boldsymbol{\sigma}_{j}\right) \sum_{a=1}^{3}\left(\boldsymbol{\lambda}_{i}^{a} \cdot \boldsymbol{\lambda}_{j}^{a}\right),
\end{align}
\begin{align}
	V_{K}\left(\boldsymbol{r}_{i j}\right) =&\frac{g_{c h}^{2}}{4 \pi} \frac{m_{K}^{2}}{12 m_{i} m_{j}} \frac{\Lambda_{K}^{2}}{\Lambda_{K}^{2}-m_{K}^{2}} m_{K}\left[Y\left(m_{K} \boldsymbol{r}_{i j}\right)\right. \nonumber \\
	&\left.-\frac{\Lambda_{K}^{3}}{m_{K}^{3}} Y\left(\Lambda_{K} \boldsymbol{r}_{i j}\right)\right]\left(\boldsymbol{\sigma}_{i} \cdot \boldsymbol{\sigma}_{j}\right) \sum_{a=4}^{7}\left(\boldsymbol{\lambda}_{i}^{a} \cdot \boldsymbol{\lambda}_{j}^{a}\right),
\end{align}
\begin{align}
	V_{\eta}\left(\boldsymbol{r}_{i j}\right) =&\frac{g_{c h}^{2}}{4 \pi} \frac{m_{\eta}^{2}}{12 m_{i} m_{j}} \frac{\Lambda_{\eta}^{2}}{\Lambda_{\eta}^{2}-m_{\eta}^{2}} m_{\eta}\left[Y\left(m_{\eta} \boldsymbol{r}_{i j}\right)\right. \nonumber \\
	&\left.-\frac{\Lambda_{\eta}^{3}}{m_{\eta}^{3}} Y\left(\Lambda_{\eta} \boldsymbol{r}_{i j}\right)\right]\left(\boldsymbol{\sigma}_{i} \cdot \boldsymbol{\sigma}_{j}\right)\left[\cos \theta_{p}\left(\boldsymbol{\lambda}_{i}^{8} \cdot \boldsymbol{\lambda}_{j}^{8}\right)\right.  \nonumber \\
	&\left.-\sin \theta_{p}\left(\boldsymbol{\lambda}_{i}^{0} \cdot \boldsymbol{\lambda}_{j}^{0}\right)\right],
\end{align}
where $Y(x) = e^{-x}/x$ is the standard Yukawa function.
The physical $\eta$ meson is considered by introducing the angle $\theta_{p}$ instead of the octet one. The $\boldsymbol{\lambda}^a$ are the SU(3) flavor Gell-Mann matrices.
The values of $m_\pi$, $m_k$ and $m_\eta$ are the masses of the SU(3) Goldstone bosons, which adopt the experimental values~\cite{Workman:2022ynf}.
The chiral coupling constant $g_{ch}$, is determined from the $\pi N N$ coupling constant through
\begin{align}
	\frac{g_{c h}^{2}}{4 \pi} & = \left(\frac{3}{5}\right)^{2} \frac{g_{\pi N N}^{2}}{4 \pi} \frac{m_{u, d}^{2}}{m_{N}^{2}}.
\end{align}
Assuming that flavor SU(3) is an exact symmetry, it will only be broken by the different mass of the strange quark.
The other symbols in the above expressions have their usual meanings.
All the parameters shown in Table~\ref{parameters} are fixed by masses of the ground-state baryons and mesons.
Table~\ref{hadrons} shows the masses of the baryons and mesons used in this work.
Since it is very difficult to fit well all ground-state hadrons with limited parameters, we give priority to fitting lighter baryons and mesons when setting parameters.
As a result, the mass gaps between theoretical and experimental values of heavier baryons and mesons are larger.
\begin{table}[ht]
	\caption{\label{parameters}Model parameters used in this work:
		$m_{\pi} = 0.7$, $m_{K} = 2.51$, $m_{\eta} = 2.77$, $\Lambda_{\pi} = 4.2$, $\Lambda_{K} = 5.2$, $\Lambda_{\eta} = 5.2$ fm$^{-1}$, $g_{ch}^2/(4\pi)$ = 0.54.}
	\begin{tabular}{cccccc}
		\hline\hline
		~~~~$b$~~~~ & ~~$m_{q}$~~ & ~~~$m_{s}$~~~  & ~~~$V_{0_{qq}}$~~~~&~~~$V_{0_{q\bar{q}}}$~~~~& ~~~$ a_c$~~~   \\
		(fm)        & (MeV)       & (MeV)          & (fm$^{-2}$)        & (fm$^{-2}$)             & ~(MeV\,fm$^{-2}$)~ \\
		0.518       & 313         & 573            &  $-$1.288          &  $-$0.201               &  58.03 \\
		\hline
		$\alpha_{s_{qq}}$   & $\alpha_{s_{qs}}$ & $\alpha_{s_{ss}}$ & $\alpha_{s_{q\bar{q}}}$ & $\alpha_{s_{s\bar{q}}}$   \\
		0.565               & 0.524             & 0.451             & 1.793                   &   1.783              \\
		\hline\hline	
	\end{tabular}
\end{table}
\begin{table}[ht]
	\caption{\label{hadrons}The masses of the baryons and mesons. Experimental values are taken from the Particle Data Group (PDG)~\cite{Workman:2022ynf} (in MeV).}
	\begin{tabular}{c c c c}
		\hline \hline
		Hadron & ~~~~~$I(J^P)$~~~~~  & ~~~~~~$M^{\text{Exp}}$~~~~~~ & ~$M^{\text{Theo}}$~  \\
		\hline
		$N$        & $1/2(1/2^+)$ & 939   & 939    \\
		$\Delta$   & $3/2(3/2^+)$ & 1232  & 1232   \\
		$\Lambda$  & $0(1/2^+)$   & 1115  & 1123   \\
		$\Sigma$   & $1(1/2^+)$   & 1189  & 1238   \\
		$\Sigma^*$ & $1(3/2^+)$   & 1385  & 1360   \\
		$\Xi$      & $1/2(1/2^+)$ & 1318  & 1374   \\
		$\Xi^*$    & $1/2(3/2^+)$ & 1535  & 1496   \\
		$\eta$     & $0(0^-)$     & 582   & 284   \\
		$\omega$   & $0(1^-)$     & 782   & 842   \\
		$K$        &$1/2(1^-)$    & 495   & 495   \\
		$K^*$      &$1/2(1^-)$    & 892   & 892   \\
		$\pi$      & $1(0^-)$     & 139   & 139   \\
		$\rho$     & $1(1^-)$     & 770   & 890   \\	
		\hline\hline
	\end{tabular}
	\label{hadrons}
\end{table}

In the QDCSM, quark delocalization was introduced to enlarge the model variational space to take into account the mutual distortion or the internal excitations of nucleons in the course of interaction.
It is realized by specifying the single-particle orbital wave function of the QDCSM as a linear combination of left and right Gaussians, the single-particle orbital wave functions used in the ordinary quark cluster model
\begin{eqnarray}
	\psi_{\alpha}(\boldsymbol {S_{i}} ,\epsilon) & = & \left(
	\phi_{\alpha}(\boldsymbol {S_{i}})
	+ \epsilon \phi_{\alpha}(-\boldsymbol {S_{i}})\right) /N(\epsilon), \nonumber \\
	\psi_{\beta}(-\boldsymbol {S_{i}} ,\epsilon) & = &
	\left(\phi_{\beta}(-\boldsymbol {S_{i}})
	+ \epsilon \phi_{\beta}(\boldsymbol {S_{i}})\right) /N(\epsilon), \nonumber \\
	N(S_{i},\epsilon) & = & \sqrt{1+\epsilon^2+2\epsilon e^{-S_i^2/4b^2}}. \label{1q}
\end{eqnarray}
It is worth noting that the mixing parameter $\epsilon$ is not an adjusted one but determined variationally by the dynamics of the multiquark system itself.
In this way, the multiquark system chooses its favorable configuration in the interacting process.
This mechanism has been used to explain the crossover transition between the hadron phase and quark-gluon plasma phase~\cite{Xu}.

\subsection{Resonating group method for bound-state and scattering process}
The resonating group method (RGM)~\cite{RGM1,RGM} and generating coordinates method~\cite{GCM1,GCM2} are used to carry out a dynamical calculation.
The main feature of the RGM for two-cluster systems is that it assumes that two clusters are frozen inside, and only considers the relative motion between the two clusters.
So the conventional ansatz for the two-cluster wave functions is
\begin{equation}
	\psi_{5q} = {\cal A }\left[[\phi_{B}\phi_{M}]^{[\sigma]IS}\otimes\chi(\boldsymbol{R})\right]^{J}, \label{5q}
\end{equation}
where the symbol ${\cal A }$ is the antisymmetrization operator, and ${\cal A} = 1-P_{14}-P_{24}-P_{34}$. $[\sigma]=[222]$ gives the total color symmetry and all other symbols have their usual meanings.
$\phi_{B}$ and $\phi_{M}$ are the $q^{3}$ and $\bar{q}q$ cluster wave functions, respectively.
From the variational principle, after variation with respect to the relative motion wave function $\chi(\boldsymbol{\mathbf{R}})=\sum_{L}\chi_{L}(\boldsymbol{\mathbf{R}})$, one obtains the RGM equation:
\begin{equation}
	\int H(\boldsymbol{\mathbf{R}},\boldsymbol{\mathbf{R'}})\chi(\boldsymbol{\mathbf{R'}})d\boldsymbol{\mathbf{R'}}=E\int N(\boldsymbol{\mathbf{R}},\boldsymbol{\mathbf{R'}})\chi(\boldsymbol{\mathbf{R'}})d\boldsymbol{\mathbf{R'}},  \label{RGM eq}
\end{equation}
where $H(\boldsymbol{\mathbf{R}},\boldsymbol{\mathbf{R'}})$ and $N(\boldsymbol{\mathbf{R}},\boldsymbol{\mathbf{R'}})$ are Hamiltonian and norm kernels.
By solving the RGM equation, we can get the energies $E$ and the wave functions.
In fact, it is not convenient to work with the RGM expressions.
Then, we expand the relative motion wave function $\chi(\boldsymbol{\mathbf{R}})$ by using a set of Gaussians with different centers
\begin{align}
	\chi(\boldsymbol{R}) =& \frac{1}{\sqrt{4 \pi}}\left(\frac{6}{5 \pi b^{2}}\right)^{3 / 4} \sum_{i,L,M} C_{i,L} \nonumber     \\
	&\cdot\int \exp \left[-\frac{3}{5 b^{2}}\left(\boldsymbol{R}-\boldsymbol{S}_{i}\right)^{2}\right] Y_{L,M}\left(\hat{\boldsymbol{S}}_{i}\right) d \Omega_{\boldsymbol{S}_{i}}
\end{align}
where $L$ is the orbital angular momentum between two clusters, and $\boldsymbol {S_{i}}$, $i=1,2,...,n$ are the generator coordinates, which are introduced to expand the relative motion wave function. By including the center-of-mass motion:
\begin{equation}
	\phi_{C} (\boldsymbol{R}_{C}) = (\frac{5}{\pi b^{2}})^{3/4}e^{-\frac{5\boldsymbol{R}^{2}_{C}}{2b^{2}}},
\end{equation}
the ansatz Eq.~(\ref{5q}) can be rewritten as
\begin{align}
	\psi_{5 q} =& \mathcal{A} \sum_{i,L} C_{i,L} \int \frac{d \Omega_{\boldsymbol{S}_{i}}}{\sqrt{4 \pi}} \prod_{\alpha=1}^{3} \phi_{\alpha}\left(\boldsymbol{S}_{i}\right) \prod_{\beta=4}^{5} \phi_{\beta}\left(-\boldsymbol{S}_{i}\right) \nonumber \\
	& \cdot\left[\left[\chi_{I_{1} S_{1}}\left(B\right) \chi_{I_{2} S_{2}}\left(M\right)\right]^{I S} Y_{LM}\left(\hat{\boldsymbol{S}}_{i}\right)\right]^{J} \nonumber \\
	& \cdot\left[\chi_{c}\left(B\right) \chi_{c}\left(M\right)\right]^{[\sigma]}, \label{5q2}
\end{align}
where $\chi_{I_{1}S_{1}}$ and $\chi_{I_{2}S_{2}}$ are the product of the flavor and spin wave functions, and $\chi_{c}$ is the color wave function.
These will be shown in detail later.
$\phi_{\alpha}(\boldsymbol{S}_{i})$ and $\phi_{\beta}(-\boldsymbol{S}_{i})$ are the single-particle orbital wave functions with different reference centers,
\begin{align}
	\phi_{\alpha}\left(\boldsymbol{S}_{i}\right) & = \left(\frac{1}{\pi b^{2}}\right)^{3 / 4} e^{-\frac{1}{2 b^{2}}\left(r_{\alpha}-\frac{2}{5} \boldsymbol{S}_{i}\right)^{2}}, \nonumber \\
	\phi_{\beta}\left(\boldsymbol{-S}_{i}\right) & = \left(\frac{1}{\pi b^{2}}\right)^{3 / 4} e^{-\frac{1}{2 b^{2}}\left(r_{\beta}+\frac{3}{5} \boldsymbol{S}_{i}\right)^{2}}.
\end{align}
With the reformulated ansatz Eq.~(\ref{5q2}), the RGM Eq.~(\ref{RGM eq}) becomes an algebraic eigenvalue equation:
\begin{equation}
	\sum_{j} C_{j}H_{i,j}= E \sum_{j} C_{j}N_{i,j},
\end{equation}
where $H_{i,j}$ and $N_{i,j}$ are the Hamiltonian matrix elements and overlaps, respectively.
By solving the generalized eigenproblem, we can obtain the energy and the corresponding wave functions of the pentaquark systems.

For a scattering problem, the relative wave function is expanded as
\begin{align}
	\chi_{L}(\mathbf{R}) & =\sum_{i} C_{i} \frac{\tilde{u}_{L}\left(\boldsymbol{R}, \boldsymbol{S}_{i}\right)}{\boldsymbol{R}} Y_{L,M}(\hat{\boldsymbol{R}}),
\end{align}
with
\begin{align}
	\tilde{u}_{L}\left(\boldsymbol{R}, \boldsymbol{S}_{i}\right) & = \left\{\begin{array}{ll}
		\alpha_{i} u_{L}\left(\boldsymbol{R}, \boldsymbol{S}_{i}\right), & \boldsymbol{R} \leq \boldsymbol{R}_{C} \\
		{\left[h_{L}^{-}(\boldsymbol{k}, \boldsymbol{R})-s_{i} h_{L}^{+}(\boldsymbol{k}, \boldsymbol{R})\right] R_{A B},} & \boldsymbol{R} \geq \boldsymbol{R}_{C}
	\end{array}\right.
\end{align}
where
\begin{align}
	u_{L}\left(\boldsymbol{R}, \boldsymbol{S}_{i}\right)= & \sqrt{4 \pi}\left(\frac{6}{5 \pi b^{2}}\right)^{3 / 4} \mathbf{R} e^{-\frac{3}{5 b^{2}}\left(\boldsymbol{R}-\boldsymbol{S}_{i}\right)^{2}} \nonumber \\
	& \cdot i^{L} j_{L}\left(-i \frac{6}{5 b^{2}} S_{i}\right).
\end{align}

$h^{\pm}_L$ are the $L$th spherical Hankel functions, $k$ is the momentum of the relative motion with $k=\sqrt{2 \mu E_\text{{ie}}}$, $\mu$ is the reduced mass of two hadrons of the open channel, $E_{\text{ie}}$ is the incident energy of the relevant open channels, which can be written as $E_{\text{ie}} = E_\text{{total}} - E_\text{{th}}$, where $E_\text{{total}}$ denotes the total energy, and $E_\text{{th}}$ represents the threshold of the open channel.
$R_C$ is a cutoff radius beyond which all the strong interaction can be disregarded.
Additionally, $\alpha_i$ and $s_i$ are complex parameters that are determined by the smoothness condition at $R = R_C$ and $C_i$ satisfy $\sum_i C_i = 1$. After performing the variational procedure, a $L$th partial-wave equation for the scattering problem can be deduced as
\begin{align}
	\sum_j \mathcal{L}_{i j}^L C_j &= \mathcal{M}_i^L(i=0,1, \ldots, n-1),
\end{align}
with
\begin{align}
	\mathcal{L}_{i j}^L&=\mathcal{K}_{i j}^L-\mathcal{K}_{i 0}^L-\mathcal{K}_{0 j}^L+\mathcal{K}_{00}^L, \nonumber \\
	\mathcal{M}_i^L&=\mathcal{K}_{00}^L-\mathcal{K}_{i 0}^L,
\end{align}
and
\begin{align}
	\mathcal{K}_{i j}^L= & \left\langle\hat{\phi}_A \hat{\phi}_B \frac{\tilde{u}_L\left(\boldsymbol{R}^{\prime}, \boldsymbol{S}_i\right)}{\boldsymbol{R}^{\prime}} Y_{L,M}\left(\boldsymbol{R}^{\prime}\right)|H-E|\right. \nonumber \\
	& \left.\cdot \mathcal{A}\left[\hat{\phi}_A \hat{\phi}_B \frac{\tilde{u}_L\left(\boldsymbol{R}, \boldsymbol{S}_j\right)}{\boldsymbol{R}} Y_{L,M}(\boldsymbol{R})\right]\right\rangle .
\end{align}
By solving Eq.~(A11), we can obtain the expansion coefficients $C_i$, then the $S$-matrix element $S_L$ and the phase shifts $\delta_L$ are given by
\begin{align}
	S_L&=e^{2 i \delta_L}=\sum_{i} C_i s_i.
\end{align}

Resonances are unstable particles usually observed as bell-shaped structures in scattering cross sections of their open channels.
For a simple narrow resonance, its fundamental properties correspond to the visible cross section features: mass $M$ is at the peak position, and decay width $\Gamma$ is the half-width of the bell shape.
The cross section $\sigma_{L}$ and the scattering phase shifts $\delta_{L}$ have relations
\begin{align}
	\sigma_L&=\frac{4 \pi}{k^2}(2 L+1) \sin ^2 \delta_L.
\end{align}
Therefore, resonances can also usually be observed in the scattering phase shift, where the phase shift of the scattering channels rises through $\pi/2$ at a resonance mass.
We can obtain a resonance mass at the position of the phase shift of $\pi/2$.
The decay width is the mass difference between the phase shift of $3\pi/4$ and $\pi/4$.

\subsection{Wave function of $\boldsymbol{qss\bar{q}q}$ system}

For the spin wave function, we first construct the spin wave functions of the $q^{3}$ and $\bar{q}q$ clusters with SU(2) algebra, and then the total spin wave function of the pentaquark system is obtained by coupling the spin wave functions of two clusters together.
The spin wave functions of the $q^{3}$ and $\bar{q}q$ clusters are Eq.~(\ref{q3s}) and Eq.~(\ref{q2s}), respectively.
\begin{align}
	\label{q3s}
	\chi_{\frac{1}{2}, \frac{1}{2}}^{\sigma}(3) & = 
	\,\big|
	\begin{tabular}{|c|c|}
		\hline
		1 & 2 \\
		\hline
		3 \\
		\cline{1-1}
	\end{tabular} \,\,
	\begin{tabular}{|c|c|}
		\hline
		$\alpha$ & $\alpha$ \\
		\hline
		$\beta$ \\
		\cline{1-1}
	\end{tabular}
	\,\big > = \frac{1}{\sqrt{6}}(2 \alpha \alpha \beta-\alpha \beta \alpha-\beta \alpha \alpha), \nonumber \\
	\chi_{\frac{1}{2}, \frac{1}{2}}^{\sigma \prime}(3) & = 
	\,\big|
	\begin{tabular}{|c|c|}
		\hline
		1 & 3 \\
		\hline
		2 \\
		\cline{1-1}
	\end{tabular} \,\,
	\begin{tabular}{|c|c|}
		\hline
		$\alpha$ & $\alpha$ \\
		\hline
		$\beta$ \\
		\cline{1-1}
	\end{tabular}
	\,\big > = \frac{1}{\sqrt{2}}(\alpha \beta \alpha-\beta \alpha \alpha), \nonumber \\
	\chi_{\frac{1}{2},-\frac{1}{2}}^{\sigma}(3) & = 
	\,\big|
	\begin{tabular}{|c|c|}
		\hline
		1 & 2 \\
		\hline
		3 \\
		\cline{1-1}
	\end{tabular} \,\,
	\begin{tabular}{|c|c|}
		\hline
		$\alpha$ & $\beta$ \\
		\hline
		$\beta$ \\
		\cline{1-1}
	\end{tabular}
	\,\big > = \frac{1}{\sqrt{6}}(\alpha \beta \beta+\beta \alpha \beta-2 \beta \beta \alpha), \nonumber \\
	\chi_{\frac{1}{2},-\frac{1}{2}}^{\sigma \prime}(3) & = 
	\,\big|
	\begin{tabular}{|c|c|}
		\hline
		1 & 3 \\
		\hline
		2 \\
		\cline{1-1}
	\end{tabular} \,\,
	\begin{tabular}{|c|c|}
		\hline
		$\alpha$ & $\beta$ \\
		\hline
		$\beta$ \\
		\cline{1-1}
	\end{tabular}
	\,\big > = \frac{1}{\sqrt{2}}(\alpha \beta \beta-\beta \alpha \beta),  \nonumber \\	
	\chi_{\frac{3}{2}, \frac{3}{2}}^{\sigma}(3) &=  
	\big|
	\begin{tabular}{|c|c|c|}
		\hline
		1 & 2 & 3 \\
		\hline
	\end{tabular} \,\,
	\begin{tabular}{|c|c|c|}
		\hline
		$\alpha$ & $\alpha$ & $\alpha$ \\
		\hline
	\end{tabular}
	\,\big > = \alpha \alpha \alpha,  \nonumber \\	
	\chi_{\frac{3}{2}, \frac{1}{2}}^{\sigma}(3) & = 
	\big|
	\begin{tabular}{|c|c|c|}
		\hline
		1 & 2 & 3 \\
		\hline
	\end{tabular} \,\,
	\begin{tabular}{|c|c|c|}
		\hline
		$\alpha$ & $\alpha$ & $\beta$ \\
		\hline
	\end{tabular}
	\,\big > =  \frac{1}{\sqrt{3}}(\alpha \alpha \beta+\alpha \beta \alpha+\beta \alpha \alpha),  \nonumber \\
	\chi_{\frac{3}{2}, -\frac{1}{2}}^{\sigma}(3) & =  
	\big|
	\begin{tabular}{|c|c|c|}
		\hline
		1 & 2 & 3 \\
		\hline
	\end{tabular} \,\,
	\begin{tabular}{|c|c|c|}
		\hline
		$\alpha$ & $\beta$ & $\beta$ \\
		\hline
	\end{tabular}
	\,\big > =  \frac{1}{\sqrt{3}}(\alpha \beta \beta+\beta \alpha \beta+\beta \beta \alpha),  \nonumber \\
	\chi_{\frac{3}{2}, -\frac{3}{2}}^{\sigma}(3) &= 
	\big|
	\begin{tabular}{|c|c|c|}
		\hline
		1 & 2 & 3 \\
		\hline
	\end{tabular} \,\,
	\begin{tabular}{|c|c|c|}
		\hline
		$\beta$ & $\beta$ & $\beta$ \\
		\hline
	\end{tabular}
	\,\big > = \beta \beta \beta.
\end{align}
\begin{align}
	\label{q2s}
	\chi_{0,0}^{\sigma}(2) & =  
	\,\big|
	\begin{tabular}{|c|}
		\hline
		1  \\
		\hline
		2 \\
		\hline
	\end{tabular} \,\,
	\begin{tabular}{|c|}
		\hline
		$\alpha$  \\
		\hline
		$\beta$ \\
		\hline
	\end{tabular}
	\,\big > =  \frac{1}{\sqrt{2}}(\alpha \beta-\beta \alpha),   \nonumber \\
	\chi_{1,1}^{\sigma}(2) &  =
	\,\big|
	\begin{tabular}{|c|c|}
		\hline
		1 & 2 \\
		\hline
	\end{tabular} \,\,
	\begin{tabular}{|c|c|}
		\hline
		$\alpha$ & $\alpha$ \\
		\hline
	\end{tabular}
	\,\big > =  \alpha  \alpha,  \nonumber \\
	\chi_{1,0}^{\sigma}(2) &  =
	\,\big|
	\begin{tabular}{|c|c|}
		\hline
		1 & 2 \\
		\hline
	\end{tabular} \,\,
	\begin{tabular}{|c|c|}
		\hline
		$\alpha$ & $\beta$ \\
		\hline
	\end{tabular}
	\,\big > =  \frac{1}{\sqrt{2}}(\alpha \beta+\beta \alpha), \nonumber \\
	\chi_{1,-1}^{\sigma}(2) &  =
	\,\big|
	\begin{tabular}{|c|c|}
		\hline
		1 & 2 \\
		\hline
	\end{tabular} \,\,
	\begin{tabular}{|c|c|}
		\hline
		$\beta$ & $\beta$ \\
		\hline
	\end{tabular}
	\,\big > =  \beta \beta.
\end{align}

For pentaquark system, the total spin quantum number can be 1/2, 3/2 or 5/2.
Considering that the Hamiltonian does not contain an interaction which can distinguish the third component of the spin quantum number, so the wave function of each spin quantum number can be written as follows
\begin{align}
	\chi_{\frac{1}{2}, \frac{1}{2}}^{\sigma 1}(5) = &\chi_{\frac{1}{2}, \frac{1}{2}}^{\sigma}(3) \chi_{0,0}^{\sigma}(2), \nonumber\\
	\chi_{\frac{1}{2}, \frac{1}{2}}^{\sigma 2}(5) = &-\sqrt{\frac{2}{3}} \chi_{\frac{1}{2},-\frac{1}{2}}^{\sigma}(3) \chi_{1,1}^{\sigma}(2)+\sqrt{\frac{1}{3}} \chi_{\frac{1}{2}, \frac{1}{2}}^{\sigma}(3) \chi_{1,0}^{\sigma}(2),   \nonumber\\
	\chi_{\frac{1}{2}, \frac{1}{2}}^{\sigma 3}(5) = & \sqrt{\frac{1}{6}} \chi_{\frac{3}{2},-\frac{1}{2}}^{\sigma}(3) \chi_{1,1}^{\sigma}(2)-\sqrt{\frac{1}{3}} \chi_{\frac{3}{2}, \frac{1}{2}}^{\sigma}(3) \chi_{1,0}^{\sigma}(2) \nonumber \\
	&+\sqrt{\frac{1}{2}} \chi_{\frac{3}{2}, \frac{3}{2}}^{\sigma}(3) \chi_{1,-1}^{\sigma}(2),  \nonumber \\
	\chi_{\frac{3}{2}, \frac{3}{2}}^{\sigma 4}(5) = & \chi_{\frac{1}{2}, \frac{1}{2}}^{\sigma}(3) \chi_{1,1}^{\sigma}(2 ),  \nonumber \\
	\chi_{\frac{3}{2}, \frac{3}{2}}^{\sigma 5}(5) = & \chi_{\frac{3}{2}, \frac{3}{2}}^{\sigma}(3) \chi_{0,0}^{\sigma}(2), \nonumber \\
	\chi_{\frac{3}{2}, \frac{3}{2}}^{\sigma 6}(5) = & \sqrt{\frac{3}{5}} \chi_{\frac{3}{2}, \frac{3}{2}}^{\sigma}(3) \chi_{1,0}^{\sigma}(2)-\sqrt{\frac{2}{5}} \chi_{\frac{3}{2}, \frac{1}{2}}^{\sigma}(3) \chi_{1,1}^{\sigma}(2), \nonumber \\
	\chi_{\frac{5}{2}, \frac{5}{2}}^{\sigma 7}(5) = & \chi_{\frac{3}{2}, \frac{3}{2}}^{\sigma}(3) \chi_{1,1}^{\sigma}(2).
\end{align}

Similar to constructing spin wave functions, we first write down the flavor wave functions of the $q^{3}$ clusters, which are
\begin{align}
	\label{q3f}
	\chi_{0,0}^{f}(3) & = 
	\,\big|
	\begin{tabular}{|c|c|}
		\hline
		1 & 2 \\
		\hline
		3 \\
		\cline{1-1}
	\end{tabular} \,\,
	\begin{tabular}{|c|c|}
		\hline
		$u$ & $s$ \\
		\hline
		$d$ \\
		\cline{1-1}
	\end{tabular}
	\,\big > = \sqrt{\frac{1}{4}}(usd + sud - sdu - dsu), \nonumber \\
	\chi_{0,0}^{f \prime}(3) & = 
	\,\big|
	\begin{tabular}{|c|c|}
		\hline
		1 & 3 \\
		\hline
		2 \\
		\cline{1-1}
	\end{tabular} \,\,
	\begin{tabular}{|c|c|}
		\hline
		$u$ & $s$ \\
		\hline
		$d$ \\
		\cline{1-1}
	\end{tabular}
	\,\big >  \nonumber \\
	& = \sqrt{\frac{1}{12}}(2uds - 2dus + sdu +usd - sud -dsu), \nonumber \\
	\chi_{\frac{1}{2},\frac{1}{2}}^{f}(3) & = 
	\,\big|
	\begin{tabular}{|c|c|}
		\hline
		1 & 2 \\
		\hline
		3 \\
		\cline{1-1}
	\end{tabular} \,\,
	\begin{tabular}{|c|c|}
		\hline
		$u$ & $s$ \\
		\hline
		$s$ \\
		\cline{1-1}
	\end{tabular}
	\,\big > = \sqrt{\frac{1}{6}}(uss + sus -2ssu), \nonumber \\
	\chi_{\frac{1}{2},\frac{1}{2}}^{f \prime}(3) & = 
	\,\big|
	\begin{tabular}{|c|c|}
		\hline
		1 & 3 \\
		\hline
		2 \\
		\cline{1-1}
	\end{tabular} \,\,
	\begin{tabular}{|c|c|}
		\hline
		$u$ & $s$ \\
		\hline
		$s$ \\
		\cline{1-1}
	\end{tabular}
	\,\big > = \sqrt{\frac{1}{2}}(uss - sus),  \nonumber \\	
	\chi_{\frac{1}{2},\frac{1}{2}}^{f \prime \prime}(3) & = 
	\big|
	\begin{tabular}{|c|c|c|}
		\hline
		1 & 2 & 3 \\
		\hline
	\end{tabular} \,\,
	\begin{tabular}{|c|c|c|}
		\hline
		$u$ & $s$ & $s$ \\
		\hline
	\end{tabular}
	\,\big > =  \sqrt{\frac{1}{3}}(uss + sus + ssu),  \nonumber \\
	\chi_{\frac{1}{2}, -\frac{1}{2}}^{f}(3) & = 
	\,\big|
	\begin{tabular}{|c|c|}
		\hline
		1 & 2 \\
		\hline
		3 \\
		\cline{1-1}
	\end{tabular} \,\,
	\begin{tabular}{|c|c|}
		\hline
		$d$ & $s$ \\
		\hline
		$s$ \\
		\cline{1-1}
	\end{tabular}
	\,\big > = \sqrt{\frac{1}{6}}(dss + sds -2ssd), \nonumber \\
	\chi_{\frac{1}{2}, -\frac{1}{2}}^{f \prime}(3) & = 
	\,\big|
	\begin{tabular}{|c|c|}
		\hline
		1 & 3 \\
		\hline
		2 \\
		\cline{1-1}
	\end{tabular} \,\,
	\begin{tabular}{|c|c|}
		\hline
		$d$ & $s$ \\
		\hline
		$s$ \\
		\cline{1-1}
	\end{tabular}
	\,\big > = \sqrt{\frac{1}{2}}(dss - sds),  \nonumber \\	
	\chi_{\frac{1}{2}, -\frac{1}{2}}^{f \prime \prime}(3) & = 
	\big|
	\begin{tabular}{|c|c|c|}
		\hline
		1 & 2 & 3 \\
		\hline
	\end{tabular} \,\,
	\begin{tabular}{|c|c|c|}
		\hline
		$d$ & $s$ & $s$ \\
		\hline
	\end{tabular}
	\,\big > =  \sqrt{\frac{1}{3}}(dss + sds + ssd),  \nonumber \\
	\chi_{1,1}^{f}(3) & = 
	\,\big|
	\begin{tabular}{|c|c|}
		\hline
		1 & 2 \\
		\hline
		3 \\
		\cline{1-1}
	\end{tabular} \,\,
	\begin{tabular}{|c|c|}
		\hline
		$u$ & $u$ \\
		\hline
		$s$ \\
		\cline{1-1}
	\end{tabular}
	\,\big > = \sqrt{\frac{1}{6}}(2uus -usu - suu), \nonumber \\
	\chi_{1,1}^{f \prime}(3) & = 
	\,\big|
	\begin{tabular}{|c|c|}
		\hline
		1 & 3 \\
		\hline
		2 \\
		\cline{1-1}
	\end{tabular} \,\,
	\begin{tabular}{|c|c|}
		\hline
		$u$ & $u$ \\
		\hline
		$s$ \\
		\cline{1-1}
	\end{tabular}
	\,\big > = \sqrt{\frac{1}{2}}(usu - suu), \nonumber \\
	\chi_{1,1}^{f \prime \prime}(3) & = 
	\big|
	\begin{tabular}{|c|c|c|}
		\hline
		1 & 2 & 3 \\
		\hline
	\end{tabular} \,\,
	\begin{tabular}{|c|c|c|}
		\hline
		$u$ & $u$ & $s$ \\
		\hline
	\end{tabular}
	\,\big >  =  \sqrt{\frac{1}{3}}(uus + usu + suu),  \nonumber \\
	\chi_{1,0}^{f}(3) & = 
	\,\big|
	\begin{tabular}{|c|c|}
		\hline
		1 & 2 \\
		\hline
		3 \\
		\cline{1-1}
	\end{tabular} \,\,
	\begin{tabular}{|c|c|}
		\hline
		$u$ & $d$ \\
		\hline
		$s$ \\
		\cline{1-1}
	\end{tabular}
	\,\big >  \nonumber \\
	& = \sqrt{\frac{1}{12}}(2uds + 2dus - sdu -usd - sud - dsu), \nonumber \\
	\chi_{1,0}^{f \prime}(3) & = 
	\,\big|
	\begin{tabular}{|c|c|}
		\hline
		1 & 3 \\
		\hline
		2 \\
		\cline{1-1}
	\end{tabular} \,\,
	\begin{tabular}{|c|c|}
		\hline
		$u$ & $d$ \\
		\hline
		$s$ \\
		\cline{1-1}
	\end{tabular}
	\,\big > = \sqrt{\frac{1}{4}}(usd + dsu - sdu - sud), \nonumber \\	
	\chi_{1, 0}^{f \prime \prime}(3) & = 
	\big|
	\begin{tabular}{|c|c|c|}
		\hline
		1 & 2 & 3 \\
		\hline
	\end{tabular} \,\,
	\begin{tabular}{|c|c|c|}
		\hline
		$u$ & $d$ & $s$ \\
		\hline
	\end{tabular}
	\,\big >   \nonumber \\
	&=  \sqrt{\frac{1}{6}}(uds + dus + sdu + usd + sud +dsu),  \nonumber \\
	\chi_{1,-1}^{f}(3) & = 
	\,\big|
	\begin{tabular}{|c|c|}
		\hline
		1 & 2 \\
		\hline
		3 \\
		\cline{1-1}
	\end{tabular} \,\,
	\begin{tabular}{|c|c|}
		\hline
		$d$ & $d$ \\
		\hline
		$s$ \\
		\cline{1-1}
	\end{tabular}
	\,\big > = \sqrt{\frac{1}{6}}(2dds -dsd - sdd), \nonumber \\
	\chi_{1,-1}^{f \prime}(3) & = 
	\,\big|
	\begin{tabular}{|c|c|}
		\hline
		1 & 3 \\
		\hline
		2 \\
		\cline{1-1}
	\end{tabular} \,\,
	\begin{tabular}{|c|c|}
		\hline
		$d$ & $d$ \\
		\hline
		$s$ \\
		\cline{1-1}
	\end{tabular}
	\,\big > = \sqrt{\frac{1}{2}}(dsd - sdd), \nonumber \\
	\chi_{1,1}^{f \prime \prime}(3) & = 
	\big|
	\begin{tabular}{|c|c|c|}
		\hline
		1 & 2 & 3 \\
		\hline
	\end{tabular} \,\,
	\begin{tabular}{|c|c|c|}
		\hline
		$d$ & $d$ & $s$ \\
		\hline
	\end{tabular}
	\,\big >  =  \sqrt{\frac{1}{3}}(dds + dsd + sdd).
\end{align}

Then, the flavor wave functions of $\bar{q}q$ clusters are
\begin{align}
	\chi_{0,0}^{f}(2) & = \sqrt{\frac{1}{2}}(\bar{d} d+\bar{u} u), \nonumber \\
	\chi_{\frac{1}{2}, \frac{1}{2}}^{f}(2) & = \bar{d} s, \nonumber \\
	\chi_{\frac{1}{2},-\frac{1}{2}}^{f}(2) & = -\bar{u} s, \nonumber \\
	\chi_{1,1}^{f}(2) & = \bar{d} u, \nonumber \\
	\chi_{1,0}^{f}(2) & = \sqrt{\frac{1}{2}}(\bar{d} d-\bar{u} u), \nonumber \\
	\chi_{1,-1}^{f}(2) & = -\bar{u} d.
\end{align}

As for the flavor degree of freedom, the isospin $I$ of pentaquark systems we investigated in this work is $I=1/2$.
The flavor wave functions of pentaquark systems can be expressed as
\begin{align}
	\chi_{\frac{1}{2}, \frac{1}{2}}^{f 1}(5) & = \chi_{0,0}^{f}(3) \chi_{\frac{1}{2}, \frac{1}{2}}^{f}(2),  \nonumber \\
	\chi_{\frac{1}{2}, \frac{1}{2}}^{f 2}(5) & = \chi_{\frac{1}{2}, \frac{1}{2}}^{f}(3) \chi_{0,0}^{f}(2),  \nonumber \\
	\chi_{\frac{1}{2}, \frac{1}{2}}^{f 3}(5) & = -\sqrt{\frac{2}{3}} \chi_{\frac{1}{2}, -\frac{1}{2}}^{f}(3) \chi_{1,1}^{f}(2) + \sqrt{\frac{1}{3}} \chi_{\frac{1}{2}, \frac{1}{2}}^{f}(3) \chi_{1,0}^{f}(2), \nonumber \\
	\chi_{\frac{1}{2}, \frac{1}{2}}^{f 4}(5) & = \sqrt{\frac{2}{3}} \chi_{1,1}^{f}(3) \chi_{\frac{1}{2}, -\frac{1}{2}}^{f}(2) - \sqrt{\frac{1}{3}} \chi_{1,0}^{f}(3) \chi_{\frac{1}{2}, \frac{1}{2}}^{f}(2).
\end{align}

Studies show that color screening is an effective description of the hidden-color channel coupling~\cite{ChenLZ,Huang:2011kf}.
Therefore, we only consider color-singlet channel (two clusters are color-singlet), which can be obtained by $1 \otimes 1$:
\begin{align}
	\chi^{c} =& \frac{1}{\sqrt{6}}(r g b-r b g+g b r-g r b+b r g-b g r)\nonumber \\
	&\cdot\frac{1}{\sqrt{3}}(\bar{r}r+\bar{g}g+\bar{b}b).
\end{align}

Finally, we can acquire the total wave functions by combining the wave functions of the orbital, spin, flavor and color parts together according to the quantum numbers of the pentaquark systems.

\section{The results and discussions}
In the present calculation, we systematically investigate the $S$-wave $qss\bar{q}q$ ($q=u~\text{or}~d)$ pentaquark systems in the framework of the QDCSM.
The quantum numbers $I$ = 1/2, $J^P = 1/2^-, 3/2^-$, and $5/2^-$ are considered.
The effective potential of each channel is studied and presented in the Fig.~\ref{potential1}, Fig.~\ref{potential2}, and Fig.~\ref{potential3} as the first step.
In order to find out if there exists any bound state, we carry out a dynamic bound-state calculation of both single-channel and channel coupling.
The numerical results are listed in Table~\ref{energy1}, Table~\ref{energy2}, and Table~\ref{energy3}.
Moreover, to verify whether the quasibound channel form resonance state or scattering state after channel coupling, the scattering process is also studied, which can be seen in Fig.~\ref{shift051}, Fig.~\ref{shift052}, Fig.~\ref{shift15}, and Fig.~\ref{shift25}.
The summary of the obtained states is presented in Table~\ref{sum}.

\subsection{\boldmath{$J^P = \frac{1}{2}^-$} sector}

\begin{figure}[htb]
	\centering
	\includegraphics[width=8cm]{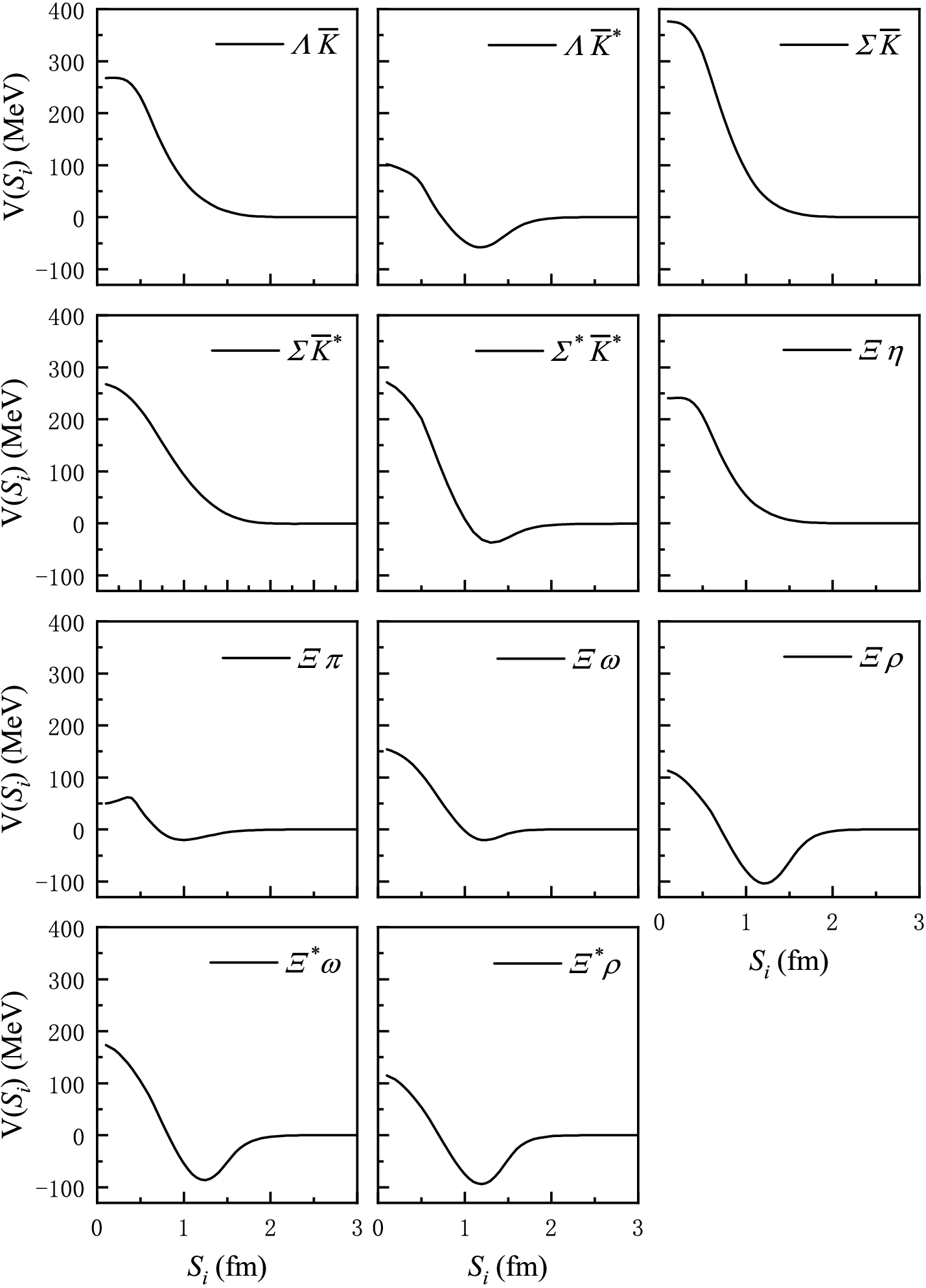}
	\caption{\label{potential1}  The effective potentials of the $qss\bar{q}q$ system with $J^P = \frac{1}{2}^-$.}
\end{figure}

The formation of states requires an attractive interaction, hence we first calculate the effective potential between baryons and mesons.
The effective potential is quantified as $V(S_i) = E(S_i) - E(\infty)$, where $S_{i}$ represents the distance between baryons and mesons.
Here, $E(S_i)$ denotes the energy of the system at the generator coordinate $S_{i}$, and $E(\infty)$ corresponds to the energy when the clusters are sufficiently far apart.
$E(S_i)$ is determined through the following expression:
\begin{align}
	E\left(S_{i}\right) & = \frac{\left\langle\Psi_{5 q}\left(S_{i}\right)|H| \Psi_{5 q}\left(S_{i}\right)\right\rangle}{\left\langle\Psi_{5 q}\left(S_{i}\right) \mid \Psi_{5 q}\left(S_{i}\right)\right\rangle},
\end{align}
where $\Psi_{5 q}(S_{i})$ represents the wave function of a certain channel, $\left\langle\Psi_{5 q}\left(S_{i}\right)|H| \Psi_{5 q}\left(S_{i}\right)\right\rangle$ and $\left\langle\Psi_{5 q}\left(S_{i}\right) \mid \Psi_{5 q}\left(S_{i}\right)\right\rangle$ are the diagonal matrix element of the Hamiltonian and the overlap, respectively.
The effective potentials of the $qss\bar{q}q$ system with $J^P = 1/2^-$ are presented in Fig.~\ref{potential1}.
There are eleven physical channels, which are $\Lambda \bar{K}$, $\Lambda \bar{K}^*$, $\Sigma \bar{K}$, $\Sigma \bar{K}^*$, $\Sigma^* \bar{K}^*$, $\Xi \eta$, $\Xi \pi$, $\Xi \omega$, $\Xi \rho$, $\Xi^* \omega$, and $\Xi^* \rho$.
Among them, the interaction of $\Xi \pi$ and $\Xi \omega$ exhibit very weak attraction at medium range.
Although $\Sigma^* \bar{K}^*$ exhibits attractive interaction, the repulsive interaction at close range is very stong.
In this case, it is difficult for $\Sigma^* \bar{K}^*$ to form a bound state.
As for $\Lambda \bar{K}^*$, $\Xi \rho$, $\Xi^* \omega$, and $\Xi^* \rho$, it is likely for these channels to form single-channel bound states.
However, $\Lambda \bar{K}$, $\Sigma \bar{K}$, $\Lambda \bar{K}^*$ and $\Xi \eta$ channels are likely to be scattering states considering their purely repulsive interactions.

To verify whether channels with attractive interactions form bound states, a dynamic calculation was conducted, and numerical results are detailed in Table~\ref{energy1}.
The first column labeled "Channel" identifies the relevant physical channels.
The second and third columns, $\chi^{f_i}$ and $\chi^{\sigma_j}$, respectively represent the flavor and spin wave functions of the corresponding physical channels, which are presented in Section II C.
The fourth and fifth columns, labeled $E_{\mathrm{th}}^{\mathrm{Exp}}$ and $E_{\mathrm{th}}^{\mathrm{Theo}}$ respectively, represent the theoretical threshold and experimental threshold of the system.
The sixth column labeled $E$ displays the energy of each single channel.
The binding energies $E_\mathrm{B} = E^{\mathrm{Theo}} - E_{\mathrm{th}}^{\mathrm{Theo}}$ are recorded in the seventh column only when $E_\mathrm{B} < 0$ MeV; otherwise, "Ub" indicates that the system is unbound.
After channel coupling, the lowest energy of coupled-channel and the lowest threshold are listed in the last row.

\begin{table}[htb]
	\caption{\label{energy1} The single-channel and the coupled-channel energies of the $qss\bar{q}q$ pentaquark systems with $J^P = \frac{1}{2}^-$ (in MeV).}
	\begin{tabular}{c c c c c c c c c}
		\hline\hline
		Channel & ~~$\chi^{f_i}$~~ & ~~$\chi^{\sigma_j}$~~ & ~~$E_{\text{th}}^{\text{Exp}}$~~  & ~~$E_{\mathrm{th}}^{\text{Theo}}$~~ & ~~$E^{\text{Theo}}$~~ & ~~$E_{\text{B}}$~~    \\
		\hline
		$\Lambda \bar{K}$    & $i = 1$ & $j = 1$ & 1610 & 1613 & 1617 & Ub     \\
		$\Lambda \bar{K}^*$  & $i = 1$ & $j = 2$ & 2007 & 2010 & 2009 & $-$1   \\
		$\Sigma \bar{K}$     & $i = 4$ & $j = 1$ & 1684 & 1727 & 1731 & Ub     \\
		$\Sigma \bar{K}^*$   & $i = 4$ & $j = 2$ & 2081 & 2124 & 2128 & Ub     \\
		$\Sigma^* \bar{K}^*$ & $i = 4$ & $j = 3$ & 2277 & 2247 & 2250 & Ub     \\
		
		$\Xi \eta$           & $i = 2$ & $j = 1$ & 1900 & 1679 & 1683 & Ub     \\
		$\Xi \pi$            & $i = 3$ & $j = 1$ & 1457 & 1534 & 1537 & Ub     \\
		$\Xi \omega$         & $i = 2$ & $j = 2$ & 2100 & 2238 & 2242 & Ub     \\
		$\Xi \rho$           & $i = 3$ & $j = 2$ & 2088 & 2286 & 2266 & $-$20  \\
		$\Xi^* \omega$       & $i = 2$ & $j = 3$ & 2318 & 2360 & 2352 & $-$8   \\
		$\Xi^* \rho$         & $i = 3$ & $j = 3$ & 2306 & 2408 & 2394 & $-$14  \\
		Coupling             &         &         & 1457 & 1534 & 1536 & Ub     \\
		\hline\hline
	\end{tabular}
\end{table}

It is worth noting that only the lowest energy of each channel is presented in the table.
This is because the formation of a bound state depends on whether the lowest energy falls below the threshold.
As is listed in Table~\ref{energy1}, four channels form quasibound states in the single-channel calculation, which are $\Lambda \bar{K}^*$, $\Xi \rho$, $\Xi^* \omega$, and $\Xi^* \rho$.
$\Lambda \bar{K}^*$ form a loosely bound state with a binding energy of only $-1$ MeV.
The binding energies of $\Xi \rho$, $\Xi^* \omega$, and $\Xi^* \rho$ are $-20$ MeV, $-8$ MeV, and $-14$ MeV, respectively.
According to the effective potential, the attractions of $\Sigma^* \bar{K}^*$, $\Xi \pi$, and $\Xi \omega$ are very weak, and the single-channel calculation results for these four channels are unbound.
The attractive interaction of $\Xi \rho$, $\Xi^* \omega$, and $\Xi^* \rho$ is stronger than that of $\Lambda \bar{K}^*$, hence the binding energies of $\Xi \rho$, $\Xi^* \omega$, and $\Xi^* \rho$ are deeper.
As for channels with purely repulsive interaction, $\Lambda \bar{K}$, $\Sigma \bar{K}^*$, $\Sigma^* \bar{K}^*$ and $\Xi \eta$, they are all unbound.
The numerical results of the single-channel calculation are consistent with the analysis of interaction.

While four quasibound states have been identified, none of them corresponds to the system's lowest threshold.
In this scenario, due to interactions with other channels after channel coupling, their energies might be elevated above their respective thresholds, resulting in scattering states, or they may remain below their respective thresholds, forming resonance states.
This process will be further investigated in subsequent scattering processes.
Furthermore, if the energy of the coupled channel is lower than the lowest threshold $\Xi \pi$, a stable bound state is genuinely achieved in the current system.
After conducting the channel coupling calculation, the lowest energy of coupled-channel is 1536 MeV, which exceeds the lowest threshold $\Xi \pi$ of 1534 MeV.
Therefore, the $qss\bar{q}q$ system with $J^P = 1/2^-$ does not form a bound state.

\begin{figure}[htb]
	\centering
	\includegraphics[width=8cm]{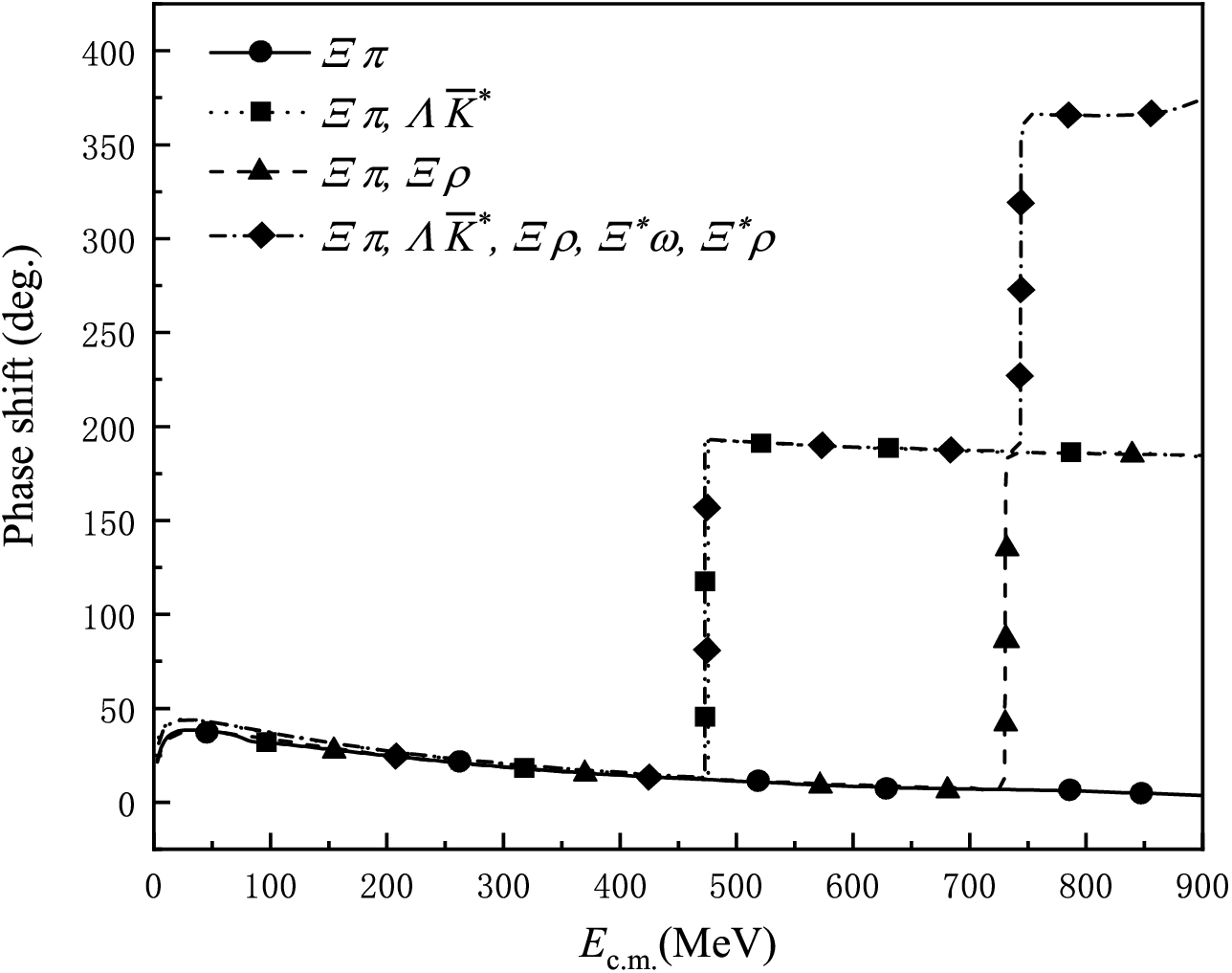}\
	\caption{\label{shift051}  The phase shifts of the open channel $\Xi \pi$ with $J^P = \frac{1}{2}^-$.}
\end{figure}

To examine whether the four quasibound channels form scattering states or resonance states, the scattering process is studied.
The details of scattering process can be found in section II B.
In Fig.~\ref{shift051}, the scattering phase shifts of the open channel $\Xi \pi$ with varying degrees of channel coupling are shown.
We separately calculate the phase shifts of the open channel $\Xi \pi$ in different coupling scenarios: (1) without coupling (solid line with circular symbols), (2) coupling the $\Xi \pi$ with a quasibound channel $\Lambda \bar{K}^*$ (dashed line with square symbols), (3) coupling the $\Xi \pi$ with a quasibound channel $\Xi \rho$ (dotted line with triangular symbols), (4) coupling the $\Xi \pi$ with four quasibound channels (dash-dotted line with diamond symbols).

As one can see, the phase shift of the open channel $\Xi \pi$ without coupling remains stable as the incident energy increases.
When the open channel $\Xi \pi$ is separately coupled with the channels $\Lambda \bar{K}^*$ and $\Xi \rho$, the phase shifts exhibit sharp increases at incident energies of 472 MeV and 733 MeV, respectively.
This indicates that the $\Lambda \bar{K}^*$ and $\Xi \rho$ are resonance states, where the resonance energy corresponds to the incident energy with a phase shift of $\pi/2$ plus the threshold of the open channel $\Xi \pi$.
The width of decay to $\Xi \pi$ corresponds to the difference in incident energies with phase shifts of $\pi/4$ and $3 \pi/4$.
After coupling the $\Xi \pi$ with the four quasibound channels, the phase shift exhibits sharp increase at an incident energy of 472 MeV and increases again at an incident energy of 740 MeV.
It indicates that the $\Lambda \bar{K}^*$ and $\Xi \rho$ form resonance states, which is consistent with the previous coupling result.
The obtained resonance energies are slightly different from that in the coupling of one quasibound channel and the open channel.
This is caused by the coupling process not being exactly the same, but this does not affect the conclusion of the resonance states.

However, after the incident energy exceeds 750 MeV, the phase shift remains stable.
No sharp increase in phase shift is shown at the incident energies corresponding to the $\Xi^* \omega$ and $\Xi^* \rho$.
We examined a series of eigenvalues obtained after channel coupling, where the lowest energy states predominantly composed of $\Lambda \bar{K}^*$ or $\Xi \rho$ remain below their respective thresholds, while those predominantly composed of $\Xi^* \omega$ or $\Xi^* \rho$ are above their corresponding thresholds.
Therefore, in the current system, there exist two resonance states, $\Lambda \bar{K}^*$ and $\Xi \rho$, while $\Xi^* \omega$ and $\Xi^* \rho$ are scattering states.

On the basis of the phase shift of $\Xi \pi$, the $\Lambda \bar{K}^*$ resonance state's theoretical resonance mass, and decay width to $\Xi \pi$ are 2006 MeV and 0.1 MeV, respectively.
The correction of the resonance mass is based on the following equation:
\begin{align}
	M^{\text{Corr}}=M^{\text{Theo}}+\sum_{n} p_{n}\left[E_{\text{th}}^{\text{Exp}}(n)-E_{\text{th}}^{\text{Theo}}(n)\right],
\end{align}
where $M^{\text{Corr}}$ and $M^{\mathrm{Theo}}$ are the corrected and theoretical mass, $p_{n}$ is the proportion of the $n$th physical channel, and $E_{\text{th}}^{\text{Exp}}(n)$ and $E_{\text{th}}^{\text{Theo}}(n)$ are the experimental and theoretical thresholds of the $n$th physical channel.
Moreover, the composition of this resonancece state is overwhelmingly predominated by the $\Lambda \bar{K}^*$.
In this scenario, the influence of other physical channels is exceedingly feeble.
Hence, based on the phase shift of the $\Xi \pi$, the corrected mass of the $\Lambda \bar{K}^*$ resonance state is 2003 MeV.
In the same way, the $\Xi \rho$ resonance state's theoretical mass, corrected mass, and decay width to the $\Xi \pi$ are 2278 MeV, 2079 MeV, and 0.2 MeV, respectively.

\begin{figure}[htb]
	\centering
	\includegraphics[width=8cm]{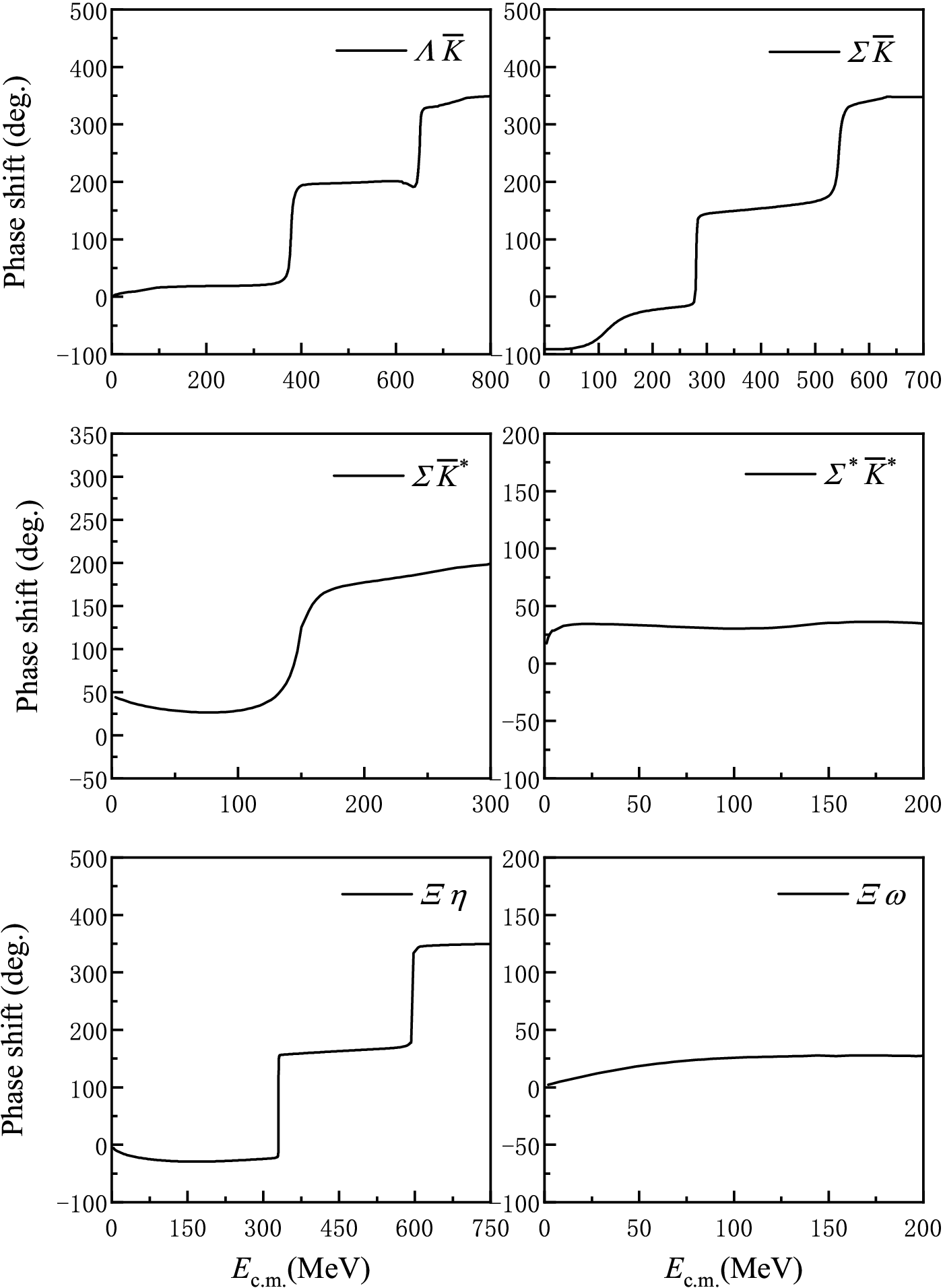}\
	\caption{\label{shift052}  The phase shifts of different open channels with $J^P = \frac{1}{2}^-$.}
\end{figure}

Further, the scattering phase shifts of other possible open channels are calculated with the coupling of the four quasibound channels and the open channel.
The results are shown in Fig.~\ref{shift052}.
The $\Lambda \bar{K}^*$ and $\Xi \rho$ resonance states can be simultaneously observed through the scattering phase shifts of the $\Lambda \bar{K}$, $\Sigma \bar{K}$, and $\Xi \eta$.
Since the threshold of $\Sigma \bar{K}^*$ being higher than the resonance energy of the $\Sigma \bar{K}$, only the $\Xi \rho$ is observed in the scattering phase shift of $\Lambda \bar{K}^*$.
The thresholds of $\Sigma^* \bar{K}^*$ and $\Xi \omega$ are higher than the energies of the two resonance states, thus no resonance state is observed.

By analyzing the scattering phase shifts, the theoretical resonance masses and the decay widths to different open channels are obtained.
The numerical results, including the corrected masses, are presented in Table~\ref{decay1}.
On this basis, the total decay widths $\Gamma_{\text{total}}$ can be estimated by $\Gamma_{\text{total}} = \Gamma_{\text{el}} +\Gamma_{\text{inel}}$, where $\Gamma_{\text{el}}$ is the elastic widths to open channels in Table~\ref{decay1}, and $\Gamma_{\text{inel}}$ is the inelastic width caused by the decay of the hadron components in resonance states.
Subsequently, we will introduce how to estimate the inelastic width of the obtained states~\cite{Ping:2008tp}.

\begin{table}[htb]
	\caption{\label{decay1} The masses and elastic widths of resonance states with the difference scattering processes (in MeV). $M^{\text{Theo}}_{\text{res}}$ and $M^{\text{Corr}}_{\text{res}}$ stand for the theoretical and corrected masses, respectively. $\Gamma_{\text{el}}$ is the elastic width of the resonance state decaying to an open channel.}
	\begin{tabular}{c c c c c c c} \hline\hline
		$J^P = 1/2^-$ & ~~~~~~~~~    & ~~$\Lambda \bar{K}$~~ & ~~$\Sigma \bar{K}$~~ & ~~$\Sigma \bar{K}^*$~~ & ~~$\Xi \eta$~~ & ~~$\Xi \pi$~~ \\
		\hline
		$\Lambda \bar{K}^*$ & $M^{\text{Theo}}$              & 1994              & 2007             & ...                & 2007       & 2006   \\
		                    & $M^{\text{Corr}}$              & 1991              & 2004             & ...                & 2004       & 2003   \\
		                    & $\Gamma_{\text{el}}$           & 8.2               & 3.7              & ...                & 0.3        & 0.1    \\
		$\Xi \rho$          & $M^{\text{Theo}}$              & 2265              & 2273             & 2270               & 2274       & 2278   \\
		                    & $M^{\text{Corr}}$              & 2066              & 2074             & 2071               & 2075       & 2079   \\
		                    & $\Gamma_{\text{el}}$           & 5.5               & 10.8             & 23.5               & 2.5        & 0.2    \\
		\hline\hline		
	\end{tabular}
\end{table}

Taking the $\Lambda \bar{K}^*$ state for example, the inelastic width $\Gamma_\text{inel}(\Lambda \bar{K}^*)$ depends on the decaying $\bar{K}^*$.
The inelastic width can be related approximately to the  $\bar{K}^*$ width $\Gamma_{f \bar{K}^*}$ = 51 MeV in free space by only accounting for the reduction in phase space available to decaying bound $\bar{K}^*$, whose mass have been reduced to roughly
\begin{align}
\label{E_B}
M_{b \bar{K}^*} \approx M_{f \bar{K}^*} - \frac{m_{\bar{K}^*}}{m_{\Lambda}+m_{\bar{K}^*}} E_B,
\end{align}
where $E_B$ is the binding energy of the bound state.
Then,
\begin{align}
\label{gamma1}
\Gamma_{b \bar{K}^*}\left(M_{b \bar{K}^*}\right) \approx \Gamma_{f \bar{K}^*} \frac{k_{b}^{2 \ell} \rho\left(M_{b \bar{K}^*}\right)}{k_{f}^{2 \ell} \rho\left(M_{f \bar{K}^*}\right)},
\end{align}
where $k$ is the pion momentum in the rest frame of the decaying $\bar{K}^*$, $\ell = 1$ is the pion angular momentum, and
\begin{align}
\label{gamma2}
\rho(M) = \pi \frac{k E_{\pi} E_{\bar{K}}}{M},
\end{align}
is the two-body decay phase space at mass $M$ when each decay product has c.m. energy $E_i$, $i = \pi, \bar{K}$.
According to the above formulae, the inelastic width $\Gamma_\text{inel}(\Lambda \bar{K}^*)$ can be obtained as $\Gamma_\text{inel}(\Lambda \bar{K}^*) = \Gamma_{b \bar{K}^*}$, which falls in the range of 47--49 MeV.
Due to the resonance energies obtained in different open channel scattering processes not being exactly the same, the binding energy in Eq.~(\ref{E_B}) has a range, leading to a range of the inelastic width as well.
Furthermore, the total decay width $\Gamma_{\text{total}}(\Lambda \bar{K}^*) = \Gamma_{\text{el}}(\Lambda \bar{K}^*) +\Gamma_{\text{inel}}(\Lambda \bar{K}^*)$ is obtained, which is 59--61 MeV.

As listed in Table~\ref{decay1}, the corrected mass of the $\Lambda \bar{K}^*$ is 1991--2004 MeV.
This state can be associated with the $\Xi(1950)$.
In fact, the experimental information regarding $\Xi(1950)$ is still not sufficiently clear.
According to the PDG's description of the mass of the $\Xi(1950)$~\cite{Workman:2022ynf}, they list experiment data reported between 1875 and 2000 MeV and suggest that there may be more than one $\Xi$ near this mass.
In addition, the decay width of the $\Lambda \bar{K}^*$ resonance is also well consistent with the experimental data of the $\Xi(1950)$.
Therefore, our calculations suggest that the $\Xi(1950)$ can be interpreted a $J^P = 1/2^-$ $\Lambda \bar{K}^*$ state.
Among other theoretical works, in Ref~\cite{Gamermann:2011mq}, $S=-2$ meson-baryon scattering in the $S$-wave is studied within a coupled channel unitary approach, explaining the $\Xi(1950)$ as a dynamically generated state with spin-parity $J^P = 1/2^-$.
The spin-parity assignment of $J^P = 1/2^-$ for the $\Xi(1950)$ is also suggested in Ref.~\cite{Oset:2010tof}.

As for the $\Xi \rho$ resonance state, the corrected mass is 2066--2079 MeV.
Based on the inelastic width $\Gamma_\text{inel}$ caused by the decaying $\rho$, the total decay width of the $\Xi \rho$ is 186--189 MeV.
Although its mass near the $\Xi(2030)$ and the $\Xi(2120)$, its decay width is much wider than the $\Xi(2030)$ and the $\Xi(2120)$.
Furthermore, its spin is also in conflict with the experimental finding of $\Xi(2030)$, which suggest that $J \ge 5/2$ ~\cite{Amsterdam-CERN-Nijmegen-Oxford:1977bvi}.
Therefore, the $\Xi \rho$ resonance state obtained in this work cannot serve as a candidate for the $\Xi(2030)$ or the $\Xi(2120)$.

\subsection{\boldmath{$J^P = \frac{3}{2}^-$} sector}

\begin{figure}[htb]
	\centering
	\includegraphics[width=8cm]{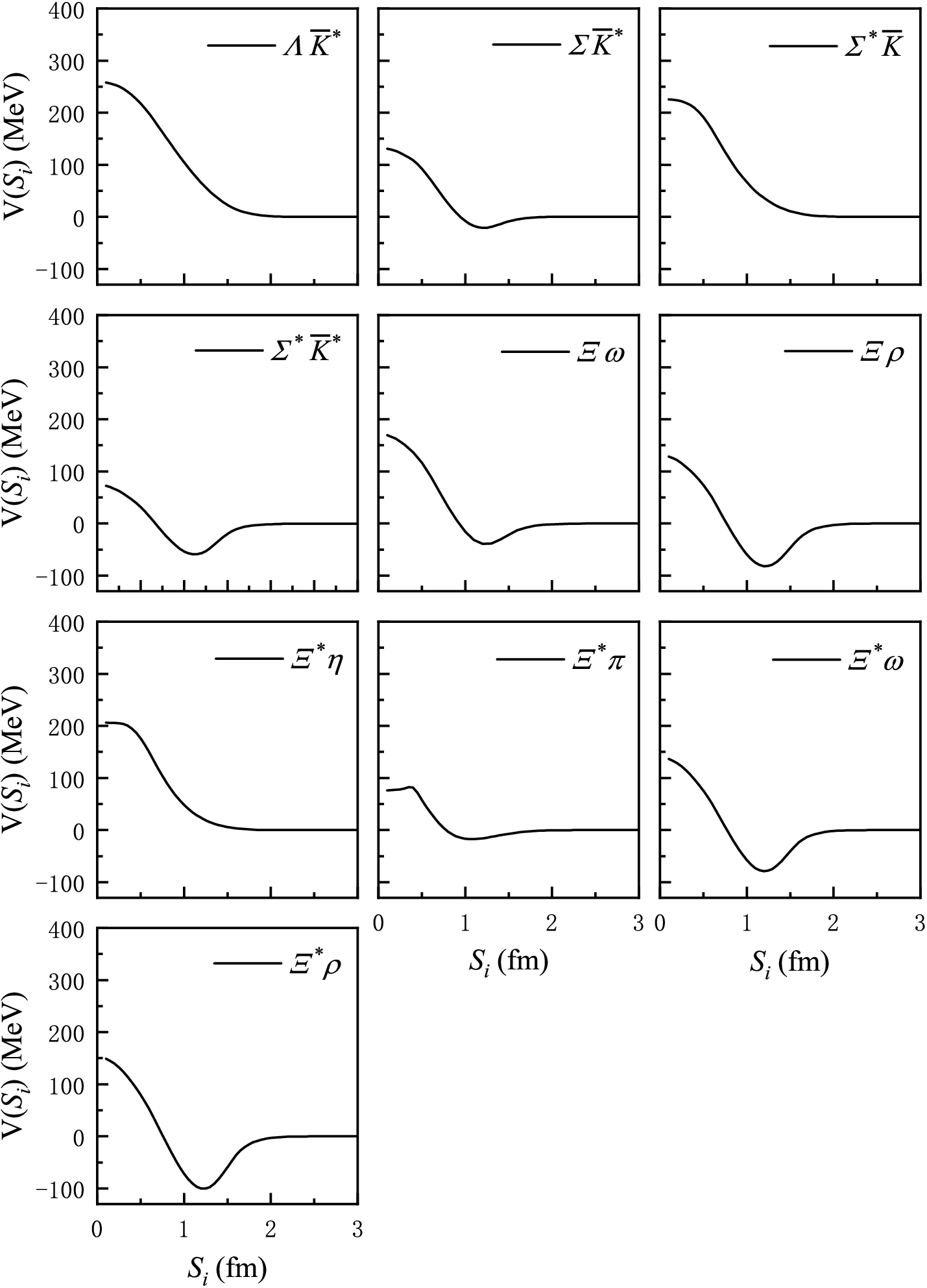}\
	\caption{\label{potential2}  The effective potentials of the $qss\bar{q}q$ system with $J^P = \frac{3}{2}^-$.}
\end{figure}

Similar to the procedure in the previous section, the effective potential of the $qss\bar{q}q$ system with $J^P = 3/2^-$ is studied.
As is shown in Fig.~\ref{potential2}, there are ten channels in this sector.
The attractive interactions of $\Xi \rho$, $\Xi^* \omega$, and $\Xi^* \rho$ channels are relatively strong, while $\Sigma \bar{K}^*$, $\Sigma^* \bar{K}^*$, $\Xi \omega$, and $\Xi^* \pi$ channels exhibit comparatively weak attractive interactions.
The interactions of $\Lambda \bar{K}^*$, $\Sigma^* \bar{K}$, and $\Xi^* \eta$ channels with $J^P = 3/2^-$ are purely repulsive, indicating that they cannot form any bound state.

The single-channel energies of the $qss\bar{q}q$ system with $J^P = 3/2^-$ and the lowest energy of coupled-channels are listed in Table~\ref{energy2}.
According to the numerical results, there are four single-channel bound states, which are $\Sigma^* \bar{K}^*$, $\Xi \rho$, $\Xi^* \omega$, and $\Xi^* \rho$.
The binding energy of $\Sigma^* \bar{K}^*$ is relatively small, at just $-4$ MeV.
This is also consistent with the weakly attractive interaction displayed by $\Sigma^* \bar{K}^*$ in the calculation of the effective potential.
The binding energies of $\Xi \rho$, $\Xi^* \omega$, and $\Xi^* \rho$ are $-15$ MeV, $-9$ MeV, and $-27$ MeV, respectively.
We will further discuss whether the four quasibound channels form resonance states or scattering states after channel coupling.
Before that, one can see that the lowest energy of the coupled-channel is 1660 MeV, still higher than the $J^P = 3/2^-$ system's lowest threshold $\Xi^* \pi$ of 1657 MeV.
Therefore, no bound state is formed in the $S$-wave $qss\bar{q}q$ system with $J^P = 3/2^-$.

\begin{table}[htb]
	\caption{\label{energy2} The single-channel and the coupled-channel energies of the $qss\bar{q}q$ pentaquark systems with $J^P = \frac{3}{2}^-$ (in MeV).}
	\begin{tabular}{c c c c c c c c c}
		\hline\hline
		Channel & ~~$\chi^{f_i}$~~ & ~~$\chi^{\sigma_j}$~~ & ~~$E_{\text{th}}^{\text{Exp}}$~~  & ~~$E_{\mathrm{th}}^{\text{Theo}}$~~ & ~~$E^{\text{Theo}}$~~ & ~~$E_{\text{B}}$~~    \\
		\hline
		$\Lambda \bar{K}^*$  & $i = 1$ & $j = 4$ & 2007 & 2010 & 2014 & Ub    \\
		$\Sigma \bar{K}^*$   & $i = 4$ & $j = 4$ & 2081 & 2124 & 2128 & Ub    \\
		$\Sigma^* \bar{K}$   & $i = 4$ & $j = 5$ & 1880 & 1850 & 1855 & Ub    \\
		$\Sigma^* \bar{K}^*$ & $i = 4$ & $j = 6$ & 2277 & 2247 & 2243 & $-$4  \\
		
		$\Xi \omega$         & $i = 2$ & $j = 4$ & 2100 & 2238 & 2241 & Ub    \\
		$\Xi \rho$           & $i = 3$ & $j = 4$ & 2088 & 2286 & 2271 & $-$15 \\
		$\Xi^* \eta$         & $i = 2$ & $j = 5$ & 2118 & 1801 & 1804 & Ub    \\
		$\Xi^* \pi$          & $i = 3$ & $j = 5$ & 1675 & 1657 & 1661 & Ub    \\
		$\Xi^* \omega$       & $i = 2$ & $j = 6$ & 2318 & 2360 & 2351 & $-$9  \\
		$\Xi^* \rho$         & $i = 3$ & $j = 6$ & 2306 & 2408 & 2381 & $-$27 \\
		Coupling             &         &         & 1675 & 1657 & 1660 & Ub    \\
		\hline\hline
	\end{tabular}
\end{table}

\begin{figure}[htb]
	\centering
	\includegraphics[width=8cm]{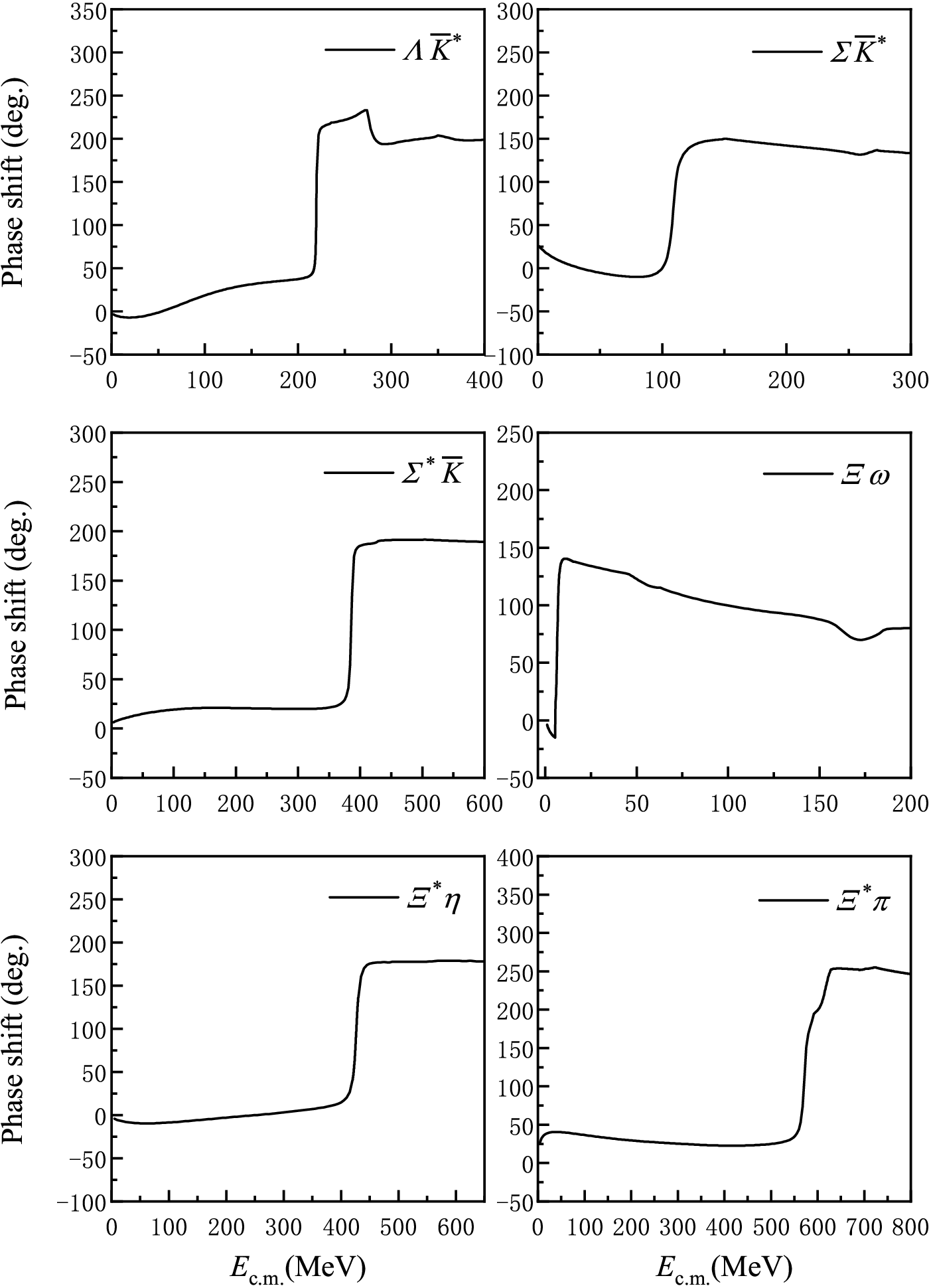}\
	\caption{\label{shift15}  The phase shifts of different open channels with $J^P = \frac{3}{2}^-$.}
\end{figure}

Four quasibound states are listed in Table~\ref{energy2}, which are $\Sigma^* \bar{K}^*$, $\Xi \rho$, $\Xi^* \omega$, and $\Xi^* \rho$.
Fig.~\ref{shift15} shows the scattering phase shifts of different open channels.
A sharp increase of phase shift is observed in the scattering process of each open channel.
Based on the resonance energies, these resonance signals correspond to the same resonance state, $\Sigma^* \bar{K}^*$.
The other three channels, $\Xi \rho$, $\Xi^* \omega$, and $\Xi^* \rho$, which form bound states in single-channel calculations, are ultimately identified as scattering states.
The widths decaying into various open channels, along with the theoretical and corrected resonance masses, are listed in the Table~\ref{decay2}.

\begin{table}[htb]
	\caption{\label{decay2} The mass and elastic width of the resonance state with the difference scattering processes (in MeV).}
	\begin{tabular}{c c c c c c c c} \hline\hline
		$J^P = 3/2^-$ & ~~~~~~~~~& ~$\Lambda \bar{K}^*$~ & ~$\Sigma \bar{K}^*$~ & ~$\Sigma^* \bar{K}$~ & ~$\Xi \omega$~ & ~$\Xi^* \eta$~ & ~$\Xi^* \pi$~\\
		\hline
		$\Sigma^* \bar{K}^*$ & $M^{\text{Theo}}$    & 2230         & 2234        & 2245           & 2239       & 2227  & 2229    \\
	                         & $M^{\text{Corr}}$    & 2260         & 2264        & 2275           & 2269       & 2257  & 2259    \\
	                         & $\Gamma_{\text{el}}$ & 4.0          & 11.0        & 5.5            & 2.1        & 8.7   & 8.5     \\
		\hline\hline	
	\end{tabular}
\end{table}

Considering the inelastic width $\Gamma_\text{inel}$ caused by the decaying $\Sigma^*$ and $\bar{K}^*$, the total decay width of the $\Sigma^* \bar{K}^*$ is given by $\Gamma_{\text{total}} = \Gamma_{\text{el}} +\Gamma_{\text{inel}}$, where $\Gamma_{\text{el}}$ can be found in Table~\ref{decay2} and $\Gamma_{\text{inel}} = \Gamma_{b \Sigma^*} + \Gamma_{b \bar{K}^*}$.
On this basis, the corrected mass and total width of the $\Sigma^* \bar{K}^*$ resonance state is obtained as 2257--2269 MeV and 116--122 MeV.
This resonance mass is consistent with the $\Xi(2250)$.
According to the PDG~\cite{Workman:2022ynf}, there are four set of experimental values on the $\Xi(2250)$ at present: \\
$~~~$Ref.~\cite{Bristol-Geneva-Heidelberg-Lausanne-QueenMaryColl-Rutherford:1986kqv}: M = 2189 $\pm$ 7 MeV, $\Gamma$ = 46 ~MeV,  \\
$~~~$Ref.~\cite{Jenkins:1983pm}: M = 2214 $\pm$ 5 MeV, $\Gamma$ = ?  \\
$~~~$Ref.~\cite{Goldwasser:1970fk}: M = 2295 $\pm$ 15 MeV, $\Gamma<$  30 MeV,  \\
$~~~$Ref.~\cite{Aachen-Berlin-CERN-London-Vienna:1969bau}: M = 2244 $\pm$ 52 MeV, $\Gamma$ = 130 $\pm$ 80 MeV. \\
Given the conflicting experimental values for decay width, there may be more than one $\Xi(2250)$.
The mass and width of the $\Sigma^* \bar{K}^*$ with $J^P =3/2^-$ ($M$ = 2257--2269 MeV, $\Gamma$ = 116--122 MeV) is consistent with the experimental values in Ref.~\cite{Aachen-Berlin-CERN-London-Vienna:1969bau} ($M = 2244 \pm 52$ MeV, $\Gamma = 130 \pm 80$ MeV).
All in all, this $J^P = 3/2^-$ $\Sigma^* \bar{K}^*$ resonance state can serve as a candidate for the two-star $\Xi(2250)$.
In other theoretical work, using the chiral unitary approach, the $\Xi(2250)$ is interpreted as a dynamically generated state with a spin-parity assignment of $J^P = 3/2^-$~\cite{Gamermann:2011mq}.

One may notice that despite $\Xi \rho$, $\Xi^* \omega$, and $\Xi^* \rho$ forming bound states in the single-channel calculation, they ultimately transition into scattering states.
Similar situations also occur in the $J^P = 1/2^-$ sector, where single-channel bound states with higher energies fail to form resonance states.
By solving the coupled-channel Schr\"{o}dinger equation, we can obtain a series of eigenenergies.
Taking the $\Sigma^* \bar{K}^*$ with $J^P = 3/2^-$ as an example, in coupled-channel calculation, $\Sigma^* \bar{K}^*$ is influenced not only by interactions from lower energies but also by interactions from higher energies.
Overall, the energy of the $\Sigma^* \bar{K}^*$ may remain stable or even be suppressed.
However, for states with higher energies, there are more eigenenergies below them than above them, making it easier for them to be elevated above their thresholds and become scattering states.

\subsection{\boldmath{$J^P = \frac{5}{2}^-$} sector}

\begin{figure}[htb]
	\centering
	\includegraphics[width=8cm]{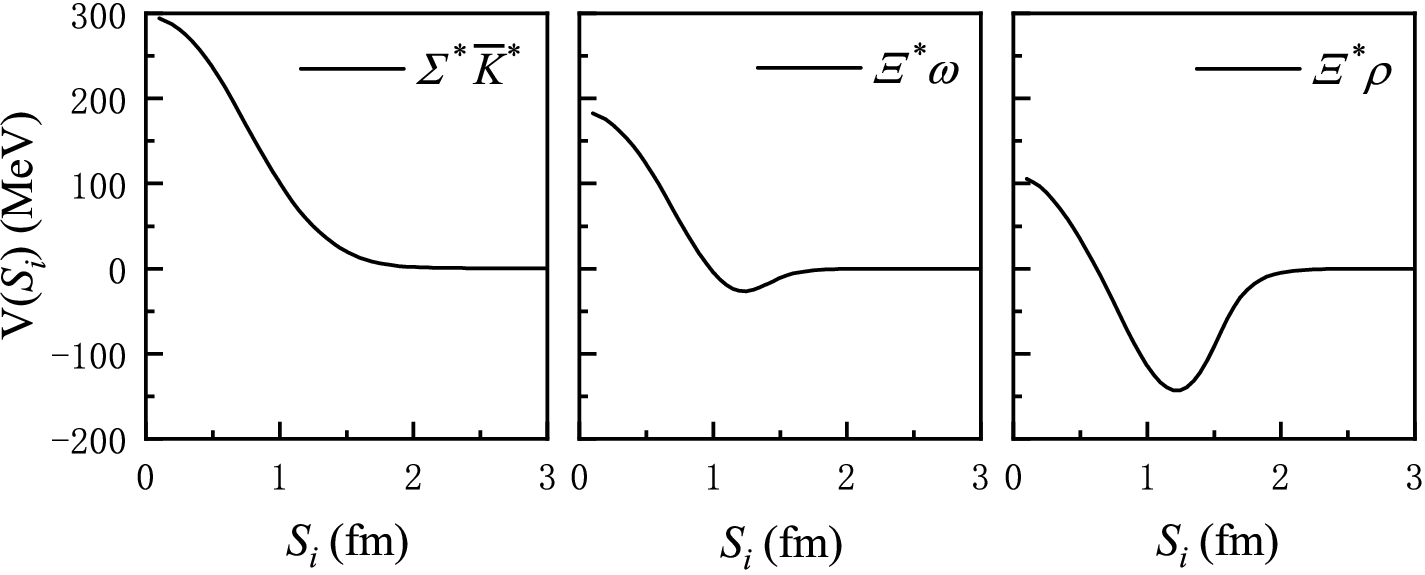}\
	\caption{\label{potential3}  The effective potentials of the $qss\bar{q}q$ system with $J^P = \frac{5}{2}^-$.}
\end{figure}

In the $J^P = 5/2^-$ sector, there are only three channels needed to be considered: $\Sigma^* \bar{K}^*$, $\Xi^* \omega$, and $\Xi^* \rho$.
As shown in Fig.~\ref{potential3}, the baryon-meson interaction of $\Sigma^* \bar{K}^*$ channel is purely repulsive.
The effective potentials of $\Xi^* \omega$ channel and $\Xi^* \rho$ channel exhibit attractions, with the attraction in the $\Xi^* \omega$ channel being very weak and the attraction in the $\Xi^* \rho$ channel being relatively strong.

Then, the results of the single-channel and coupled-channel calculations are listed in the Table~\ref{energy3}.
$\Sigma^* \bar{K}^*$ and $\Xi^* \omega$ channels are unbound in the single-channel calculations, while $\Xi^* \rho$ channel forms a quasibound state with a relatively deep binding energy of $-55$ MeV.
It is noteworthy that after the channel coupling, the lowest energy of the $qss\bar{q}q$ system with $J^P = 5/2^-$ is 2229 MeV, which is below the lowest threshold $\Sigma^* \bar{K}^*$.
Therefore, a bound state is obtained here.
According to numerical result, the predominant composition of this bound state is $\Sigma^* \bar{K}^*$, with a minor presence of $\Xi^* \omega$, while the contribution of $\Xi^* \rho$ is negligible.
On this basis, the corrected mass of this $qss\bar{q}q$ bound state with $J^P = 5/2^-$ is 2233 MeV.

\begin{table}[htb]
	\caption{\label{energy3} The single-channel and the coupled-channel energies of the $qss\bar{q}q$ pentaquark systems with $J^P = \frac{5}{2}^-$ (in MeV).}
	\begin{tabular}{c c c c c c c c c}
		\hline\hline
		Channel & ~~$\chi^{f_i}$~~ & ~~$\chi^{\sigma_j}$~~ & ~~$E_{\text{th}}^{\text{Exp}}$~~  & ~~$E_{\mathrm{th}}^{\text{Theo}}$~~ & ~~$E^{\text{Theo}}$~~ & ~~$E_{\text{B}}$~~    \\
		\hline
		$\Sigma^* \bar{K}^*$ & $i = 4$ & $j = 7$ & 2277 & 2247 & 2252 & Ub      \\
		$\Xi^* \omega$       & $i = 2$ & $j = 7$ & 2318 & 2360 & 2364 & Ub      \\
		$\Xi^* \rho$         & $i = 3$ & $j = 7$ & 2306 & 2408 & 2353 & $-$55   \\
		Coupling             &         &         & 2277 & 2247 & 2229 & $-$18   \\
		\hline\hline
	\end{tabular}
\end{table}

Since a bound state is obtained, without considering $D$-wave coupling, the total decay width equals the inelastic width.
Moreover, given that it contains two nonnegligible compositions, it should be $\Gamma_\text{inel} = \sum_{n} p_{n} \Gamma_\text{inel}(n)$, where $p_{n}$ and $\Gamma_\text{inel}(n)$ is the proportion and the inelastic width of the $n$th physical channel.
For the composition $\Sigma^* \bar{K}^*$, the inelastic width $\Gamma_\text{inel}(\Sigma^* \bar{K}^*)$ depends on the decaying $\Sigma^*$ and $\bar{K}^*$.
Hence, the inelastic width $\Gamma_\text{inel}(\Sigma^* \bar{K}^*)$ can be obtained as $\Gamma_\text{inel}(\Sigma^* \bar{K}^*) = \Gamma_{b \Sigma^*} + \Gamma_{b \bar{K}^*}$.
As for the other composition $\Xi^* \omega$, the situation is different from the previous calculation.
The dominating decay mode of $\omega$ is a three-body deacy, but Eq.~(\ref{gamma2}) is used in a two-body decay process.
However, due to the narrow decay width of $\omega$, which is only 8.7 MeV, in the calculation of Eq.~(\ref{gamma1}), the difference between the width of decaying bound $\omega$ and the width of $\omega$ in free space is very small.
We can approximately consider the width of $\omega$ in free space as the width of decaying bound $\omega$ $\Gamma_{b \omega}$.
Therefore, $\Gamma_\text{inel}(\Xi \omega) = \Gamma_{b \Xi} + \Gamma_{b \omega}$.
The total decay width of the $\Sigma^* \bar{K}^*$ bound state is 53 MeV.
As mentioned the last paragraph, the corrected mass of this bound state is 2233 MeV, which is consistent with the $\Xi(2250)$.
At the same time, the decay width of this bound state is consistent with one of the experimental values, $46 \pm 27$ MeV ~\cite{Bristol-Geneva-Heidelberg-Lausanne-QueenMaryColl-Rutherford:1986kqv}.

\begin{figure}[htb]
	\centering
	\includegraphics[width=8cm]{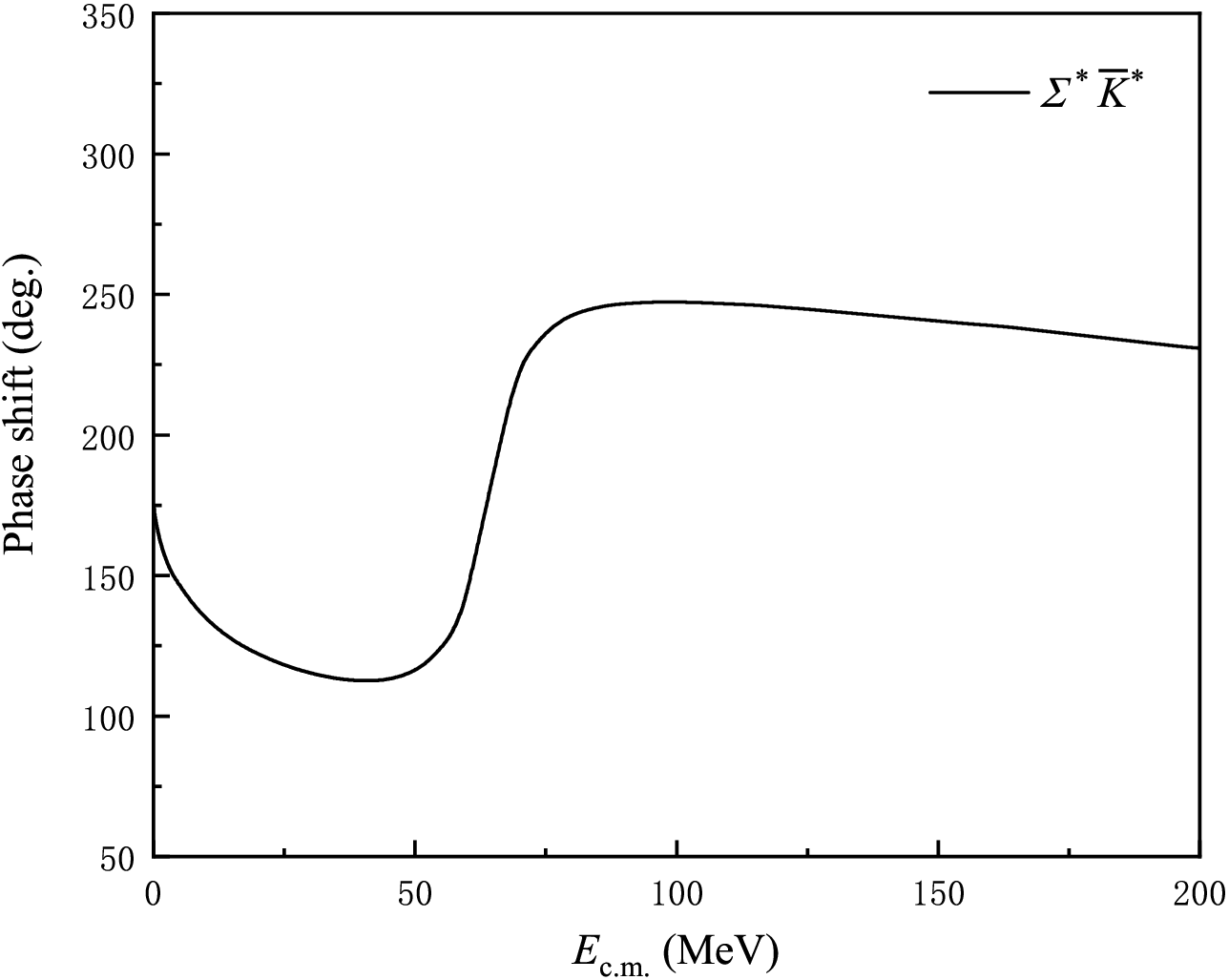}\
	\caption{\label{shift25}  The phase shift of the open channel $\Sigma^* \bar{K}^*$ with $J^P = \frac{5}{2}^-$.}
\end{figure}

Moreover, to determine the property of the quasibound state $\Xi^* \rho$, the scattering shift of the $\Sigma^* \bar{K}^*$ is shown in Fig.~\ref{shift25}.
The scattering shift increase at the incident energy around 70 MeV indicates that $\Xi^* \rho$ channel forms a resonance state.
According to the calculations, the corrected mass and the total width of the $\Xi^* \rho$ is 2240 MeV and 161 MeV, respectively.
The mass and width of the obtained $J^P = 5/2^-$  $\Xi^* \rho$ state is very consistent with the experimental value of the $\Xi(2250)$ in Ref.~\cite{Aachen-Berlin-CERN-London-Vienna:1969bau} ($M = 2244 \pm 52 ~\text{MeV},~\Gamma = 130 \pm 80$ MeV).
Additionally, one may notice that after mass correction, the energy of $\Xi^* \rho$ falls below the experimental threshold of $\Sigma^* \bar{K}^*$.
This is because in our calculations, the theoretical mass difference between the $\Sigma^* \bar{K}^*$ and $\Xi^* \rho$ is larger than the experimental mass difference between the $\Sigma^* \bar{K}^*$ and $\Xi^* \rho$, causing the $\Xi^* \rho$ with a relatively deep binding energy to still be above the theoretical threshold of the $\Sigma^* \bar{K}^*$.
But the correction mass of the $\Xi^* \rho$ is below the experimental threshold of the $\Sigma^* \bar{K}^*$.

Finally, the obtained states in this work are summarized in Table~\ref{sum}.
According to our results, the $\Lambda \bar{K}^*$ with $J^P = 1/2^-$ can serve as a candidate for the $\Xi(1950)$.
Both the $\Sigma^* \bar{K}^*$ state with $J^P = 3/2^-$ and the $\Xi^* \rho$ state with $J^P =5/2^-$ are consistent with the experimental values about the $\Xi(2250)$ given in Ref.~\cite{Aachen-Berlin-CERN-London-Vienna:1969bau}.
Moreover, the $\Sigma^* \bar{K}^*$ state with $J^P =5/2^-$ can explain the relatively narrow $\Xi(2250)$ reported in Ref.~\cite{Bristol-Geneva-Heidelberg-Lausanne-QueenMaryColl-Rutherford:1986kqv}.
On this basis, three states are obtained that can explain the $\Xi(2250)$.
This might also explain the conflicting experimental values for the width of the $\Xi(2250)$.
Additionally, we obtain a $\Xi \rho$ state with $J^P = 1/2^-$, which cannot serve as a candidate for the reported $\Xi$ resonances at present.
Therefore, we here predict a new $\Xi$ with $J^P = 1/2^-$, whose mass and width are 2066--2079 MeV and 186--189 MeV, respectively.

\begin{table}[htb]
	\caption{\label{sum} The states obtained in this work (in MeV).}
	\begin{tabular}{c c c c c c}
		\hline\hline
		$J^P$ & Main composition  & Mass & ~$\Gamma_{\text{el}}$~ & ~~~$\Gamma_{\text{inel}}$~~~ & ~~$\Gamma_{\text{total}}$~~  \\
		 \hline
		$1/2^-$  & $\Lambda \bar{K}^*$   & 1991--2004 &  12 & 47--49   & 59--61   \\
		$1/2^-$  & $\Xi \rho$            & 2066--2079 &  43 & 143--146 & 186--189 \\
		$3/2^-$  & $\Sigma^* \bar{K}^*$  & 2257--2269 &  40 & 76--82   & 116--122 \\
		$5/2^-$  & $\Sigma^* \bar{K}^*$  & 2233       & ... & 53       & 53       \\
		$5/2^-$  & $\Xi^* \rho$          & 2240       &  12 & 149      & 161       \\
		\hline\hline
	\end{tabular}
\end{table}

\section{Summary}

In this work, we investigate the $qss\bar{q}q$ system in the framework of the QDCSM.
The $S$-wave pentaquark system with $I$ = 1/2, $J^P$ = $1/2^-$, $3/2^-$, and $5/2^-$ are considered.
The effective potnetial is studied to describe the baryon-meson interactions.
The single-channel bound-state calculation and scattering process study are carried out to search and confirm resonance states.
Meanwhile, the coupled-channel dynamic bound-state calculation is performed to find bound state.
In addition, both the elastic width and the inelastic width are estimated.
Based on the current results, the conclusion can be drawn as follows.

(1) In our calculations, the $\Xi(1950)$ can be interpreted as $\Lambda \bar{K}^*$ state with $J^P = 1/2^-$.

(2) We obtain three states that match the mass of the $\Xi(2250)$.
Among them, the $\Sigma^* \bar{K}^*$ state with $J^P = 3/2^-$ and the $\Xi^* \rho$ state with $J^P =5/2^-$ are consistent with the broad width reported in Ref.~\cite{Aachen-Berlin-CERN-London-Vienna:1969bau}, while the $\Sigma^* \bar{K}^*$ state with $J^P =5/2^-$ matches the relatively narrow width reported in Ref.~\cite{Bristol-Geneva-Heidelberg-Lausanne-QueenMaryColl-Rutherford:1986kqv}.
We propose the existence of three $\Xi(2250)$ states, which might also explain the conflicting experimental values for the width.

(3) A new $\Xi$ resonance with mass and width of 2066--2079 MeV and 186--189 MeV, respectively, is predicted to exist.
This resonance is identified as the $\Xi \rho$ state with $J^P = 1/2^-$.

Compared with the $\Lambda$ and $\Sigma$ resonances, there is currently a lack of theoretical researches and experimental data on the $\Xi$ resonances.
We aim for our systematic calculations to provide valuable insights into understanding the properties of the $\Xi$ resonances and discovering new $\Xi$ resonances.
It is also necessary to understand the $\Xi$ resonances from the perspective of three-quark excited states.
Additionally, considering the mixing of three-quark and five-quark from an unquenched picture is an important aspect.

\acknowledgments{This work is supported partly by the National Natural Science Foundation of China under Contracts Nos. 11675080, No. 11775118, No. 12305087, and No. 11535005. Y. Y. is supported by the Postgraduate Research and Practice Innovation Program of Jiangsu Province under Grant No. KYCX23\underline{~}1675 and the Doctoral Dissertation Topic Funding Program under Grant No. YXXT23-027. Q. H. is supported by the Start-up Funds of Nanjing Normal University under Grant No. 184080H201B20.}

\end{document}